%% file: main.tex
\documentclass[sigconf]{acmart}
\usepackage{soul} % Allows text highlighting with line breaks

\usepackage{xcolor}
\definecolor{orange}{HTML}{ffccaa} % orange
\definecolor{blue}{HTML}{acdde8} % Light Blue
\newcommand{\higho}[1]{\sethlcolor{orange}\hl{#1}} % Yellow Highlight
\newcommand{\highb}[1]{\sethlcolor{blue}\hl{#1}} % Blue Highlight

\newcommand{\update}[1]{\textcolor{black}{#1}}
\newcommand{\up}[1]{\textcolor{black}{#1}}

\copyrightyear{2025}
\acmYear{2025}
\setcopyright{acmlicensed}\acmConference[CHI '25]{CHI Conference on Human
Factors in Computing Systems}{April 26-May 1, 2025}{Yokohama, Japan}
\acmBooktitle{CHI Conference on Human Factors in Computing Systems (CHI
'25), April 26-May 1, 2025, Yokohama, Japan}
\acmDOI{10.1145/3706598.3714035}
\acmISBN{979-8-4007-1394-1/25/04}

\begin{document}

%%
%% The "title" command has an optional parameter,
%% allowing the author to define a "short title" to be used in page headers.
\title[AI Rivalry as a Craft]{\update{AI Rivalry as a Craft: How Resisting and Embracing Generative AI Are Reshaping the Writing Profession}}

% Investigating Job crafting strategies of writing professionals who resist Generative AI.

%Tug-of-Words: Mapping Embracing and Resisting Strategies of Writing Professionals Around Generative AI Technologies

% Evolving Writing Professions: Mapping Embracing and Resisting Strategies of Writing Professionals Around Generative AI technologies

%Crafting Strategies of Writing Professionals in Response to Generative AI technologies: One year perspectives

%Crafting Strategies of Habituated Writing Professionals around Generative AI tools in work
% Crafting Strategies of Habitauated Writing Professionals around Generative AI tools in work.
%%

%TC:ignore
\author[R.Varanasi]{Rama Adithya Varanasi}
\affiliation{%
  \institution{Tandon School of Engineering \\ New York University}
  \city{NYC}
  \state{NY}
  \country{USA}
}
\author[B.Wiesenfeld]{Batia Mishan Wiesenfeld}
\affiliation{
 \institution{Stern School of Business \\ New York University}
 \city{NYC}
 \state{NY}
 \country{USA}
}
\author[O.Nov]{Oded Nov}
\affiliation{%
  \institution{Tandon School of Engineering \\ New York University}
  \city{NYC}
  \state{NY}
  \country{USA}
}




\begin{abstract}
Generative AI (GAI) technologies are disrupting professional writing, challenging traditional practices. Recent studies explore GAI adoption experiences of creative practitioners, but we know little about how these experiences evolve into established practices and how GAI resistance alters these practices. To address this gap, we conducted 25 semi-structured interviews with writing professionals who adopted and/or resisted GAI. Using the theoretical lens of Job Crafting, we identify four strategies professionals employ to reshape their roles. Writing professionals employed GAI resisting strategies to maximize human potential, reinforce professional identity, carve out a professional niche, and preserve credibility within their networks. In contrast, GAI-enabled strategies allowed writers who embraced GAI to enhance desirable workflows, minimize mundane tasks, and engage in new AI-managerial labor. These strategies amplified their collaborations with GAI while reducing their reliance on other people. We conclude by discussing implications of GAI practices on writers' identity and practices as well as crafting theory. 
\end{abstract}
\maketitle

\input{1.Introduction}

\input{2.Literature}

\input{3.Methods}
\input{4.1.expansion}

\input{4.2.delegation}

\input{5.Discussion}
\input{6.Conclusion}

\bibliographystyle{ACM-Reference-Format}
\bibliography{references}
\input{8.appendix}
\end{document}

%% file: 1.Introduction.tex
\section{Introduction}

\begin{quote}
    \textit{``If I let [G]AI do my work, It would make me miserable inside, because I really love these tasks that I do. That's the bottom line \dots The only way I can stay relevant is by opposing these tools my way.''} - P17, Screenwriter
\end{quote}
\centerline{\rule{2cm}{0.1pt}}
\begin{quote}
    \textit{``It [GAI] is helping me do my job better and make me more competitive \dots I feel individuals who are not using it [GAI] are at a serious disadvantage.''} - P4, Paralegal 
\end{quote}

\vspace{.4 cm}

In an era of rapid technological advancement, the growth of Generative AI (GAI) technologies (e.g., ChatGPT) has had an unprecedented impact on global industries, professional futures, and interpersonal relationships that few were prepared for. While the emergence of personal computers \cite{martin14}, smart devices \cite{varanasi2019}, social media \cite{Carr_Hayes_2015}, and the `first wave' of AI-integration \cite{Autor2023} have been influential in many professional and personal applications, GAI has become a direct competitor to professional writing.

Recent research has shown that new features of GAI have significant implications for all types of professional writers. Studies of their early adoption experiences associated GAI use with increased creativity, self-efficacy, and productivity \cite{Washington_2023,Doshi_Hauser_2023,Noy_Zhang_2023}. Yet during the same period, numerous writers' organizations, including the WGA and SAG-AFTRA, have expressed concerns about GAI's threat to their field \cite{rollingstone2024wgastrike}. One of the highlighted features of GAI technologies is the ability to generate creative and analytical content while engaging in natural context-sensitive conversations with humans \cite{Brynjolfsson2023}. This duality, as many professional writers have lamented, presents both opportunities and causes for concern regarding the future of work.

As GAI reshapes the literary landscape, the field of Human-Computer Interaction (HCI) has responded by focusing on the lived experiences of writers' early adoption of GAI \cite{Kobiella2024,takaffoli}. 
%For example,~\citet{Li2024}'s preliminary experimental study found that writers were willing to embrace GAI to enjoy some of its benefits. 
Among these studies, \update{there is a lack of understanding of the actions many professionals take to resist GAI and establish work practices to fight its incursion into their work. Moreover, little is known about how early experiences, whether resisting or embracing practices around GAI, evolve into habituated practices in real-world settings. Understanding the parallels between resisting and embracing practices can move discussions beyond early techno-deterministic optimism and foster equitable spaces as GAI technologies continue to mature.} 

To fill this gap, \update{we conducted a qualitative study to: (1) understand how writing professionals employ resistive strategies with GAI, and (2) compare and contrast these practices with those in similar roles who adopt embracing strategies with GAI}. We gathered data through in-depth interviews with 25 experienced professionals (avg. experience $=$ 17 years) in diverse writing fields, all of whom had been exposed to GAI for at least 12 months. Using a job crafting framework \cite{Wrzesniewski_2001}, an employee-driven process in which individuals modify their tasks and interactions to enhance their work outcomes, as our theoretical framing, we demonstrate that writing professionals who engaged in resisting practices with GAI shaped their work in distinct ways while pursuing similar goals as those who embraced GAI. Professionals who resisted GAI placed emphasis on the visibility of their work identity and human labor to their networks and end-users. In selective cases, they carved out niches for their human-driven work by appealing to exclusive communities.  

Conversely, professionals embraced GAI to enrich their workflows by delegating tedious tasks to reduce their boredom, stress, and additional emotional labor, though this sometimes extended to essential yet challenging tasks. These benefits also required reconfiguring their workflows to carve out dedicated time  for effective GAI utilization. We distill these insights into \textit{four} different resisting and embracing crafting strategies. 
%By comparing and contrasting resisting (human-driven) strategies with that of embracing (GAI-driven) strategies, our study highlights key implications for GAI \update{and creative labor, within HCI}.
We compare these strategies to show how professionals who deployed human-driven (resisting) strategies shaped both their creative identity as well as their practices, while GAI-driven (embracing) strategies were only used to craft their practices. Interestingly, professionals who deployed human-driven strategies proactively engaged in rivalry with GAI, mirroring a pattern previously observed only between professionals. This strategy differs from GAI-driven strategies that include significant invisible labor in the form of AI managerial labor. 

We argue that common metrics, like productivity and efficiency, are insufficient for assessing the value of human-driven and GAI-driven strategies, and advocate for more holistic metrics. The strategies we found offer new avenues for skill development, but also point to future skill and job displacement. To create an equitable environment for both groups, we propose fostering dialogue and collective learning. Overall, our study contributes to the development of GAI in the writing field by showing:

\begin{itemize} 
\item \up{How human-driven (resisting) strategies and GAI-driven (embracing) strategies serve different purposes in shaping writers' work, with human-driven strategies enabling the crafting of both \textit{identities} and \textit{practices}, while GAI-driven strategies enable only the crafting of \textit{practices}.} 
\item \up{How writers employ GAI-resisting strategies to develop a competitive edge using \textit{AI rivalry} --- a new phenomenon identified in our study.}
% \item \textit{Two} new forms of crafting strategies —\textit{technology crafting} and \textit{adversarial crafting}—extend existing crafting theory. 
\item \update{Ways to better integrate GAI at work to create equitable opportunities for workers engaging in both resisting and embracing techniques}.
\end{itemize}

%% file: 2.Literature.tex
\section{Related Work}
In this section, we provide an overview of HCI literature on how creative work has undergone transformation through technological innovation, highlighting key research involving workers in writing contexts. We then examine the pressing gaps within this literature by exploring the application of GAI in the creative labor and writing profession. 
%We argue for extending this preliminary research by studying the evolving practices of writing professionals, particularly those who resist GAI. 
Finally, we introduce job crafting theory \cite{Wrzesniewski_2001} as a theoretical lens for systematically investigating our research objective.

\subsection{\update{HCI and Creative Work}}

\update{Creative work is traditionally defined as a process in which ideas are developed with the intent of creating something novel or useful \cite{Hearn_2020, Harrison2022}. Craftsmanship is required to transform ideas into creations, giving creative workers autonomy to structure their processes. Craftsmanship has long required technologies (e.g., pencils, printing presses), but as more advanced technologies have become closely integrated with creative work \cite{palani24,palani2022}, HCI scholars have studied creative work's evolution, with the aim of understanding and enhancing creative processes and outcomes \cite{Hsueh24}. New tools and disciplines supporting this endeavor have emerged in the process, including computer-aided design (CAD) \cite{Robertson2007}, creativity-support tools (CST) \cite{chung2021}, and creative computing \cite{Hugill_Yang_2013}. }

\update{The integration of digital technologies into creative work has transformed creative practices in several significant ways. The first key shift was that the application of digital innovation increased the emphasis on intangible outcomes (e.g., knowledge, digital assets) over tangible ones (e.g., machinery, artifacts) \cite{Hearn_2020}. 
%The evolution of crowdsourcing platforms, such as Wikipedia and Github, is an example of how writing as creative labor branched into co-creation practices in which practitioners collaboratively produce and refine knowledge\cite{Brabham_2013}. 
This transformation reshaped value production and had significant implications for how creative labor was recognized, monetized, and distributed \cite{Ross_2008}. The second shift involved moving away from solely prioritizing creative outcomes to considering the broader dynamics of workers' creative processes \cite{Hsueh24}. For example, in creative fields like graphic design and music production, CST research has highlighted the value of supporting both conventional and unconventional practices (e.g., using digital tools as a buffer to store and develop multiple creative ideas \cite{Frinch19}) \cite{turco23, Kumaran21, palani2022}, while finding ways to extend existing practices \cite{chung22} and supporting transformed professional identities \cite{Dunn2024}. This shift has led to a reconceptualization of how creative labor is defined, understood, and valued \cite{semaan23}}.

More recently, the rise of platform-mediated creative labor on platforms like Upwork, Fiverr and Instagram has introduced new challenges and complexities to creative work. These platforms often prioritize scale, output, and client satisfaction, challenging control and autonomy for creative practitioners while blurring the boundaries between the individual and the work itself \cite{munoz22}. This dynamic has deepened the need to examine not only the activities that constitute creative labor (the ``what'') but also the identity of the person (the ``who'') performing the work. \cite{Hsueh24}. Several studies exploring the role of worker identity in relation to work practices have found that platforms often undermine workers' control over how and when they can make their identity visible, particularly for underrepresented worker communities \cite{munoz24, munoz22}. 
%For instance, \citet{semaan23} examined how TikTok's platform logic disrupts creative routines and alienates users from their intentions. 

With the advent of algorithmic oversight of gig platforms and AI-based technologies in creative work, workers' ability to control their identities and their work has become more precarious \cite{Duffy2021}. Multiple studies of online content creators have shown how algorithmic design of the platforms make it increasingly difficult for the workers to make their identities visible \cite{Duffy2023}. Similarly, platforms like Etsy have also pushed small-scale craft sellers to perform new forms of invisible labor, such as articulation work, to ensure consistent earnings \cite{razaq22}. In response to these emerging issues, HCI research is investigating how workers reconfigure their identities~\cite{choi23} and, in the process, making the underlying bottom-up efforts more explicit (e.g., algorithmic gossip used by creative workers on YouTube \cite{Bishop_2019}).
%Workers also develop explicit practices that expressed their displeasure with the algorithms while providing creative benefit to them}.

\subsection{Generative AI, HCI, and Creative Labor }

\update{GAI is the newest wave of technologies which affect work in diverse areas such as hospitality \cite{Hsu_Tan_Stantic_2024}, education \cite{Chan_Hu_2023}, manufacturing \cite{Bendoly2023}, and healthcare \cite{Small2024}, as well as in creative work \cite{23Jiang}. Recent predictions suggest that almost 80\% of the U.S. workforce could have at least 10\% of their work tasks altered by the introduction of GAI \cite{Eloundou_2023}. These significant shifts in the landscape of professional practice have given rise to two key streams of research. The first stream explores the potential of GAI to augment workers' skills and enhance their ability to perform tasks. Early experimental studies highlight beneficial worker outcomes, such as performance and productivity improvements \cite{Noy_Zhang_2023, Al2024, Dell2023, Brynjolfsson2023}. Studies have also begun to suggest that GAI tools may transform more creative aspects of work processes, such as by sparking idea generation~\cite{Epstein_2022}.} 

\update{The second stream of research cautions about several shortcomings of GAI, revealing that professionals using GAI in their work tasks could introduce bias and produce errors, degrading their outputs \cite{Miyazaki2024,Samuelson_2023, Kidd_Birhane_2023}. Moreover, by using GAI tools, workers might diminish their professional capacities, diluting human skills such as creativity and critical thinking \cite{Walczak_2023, Epstein2023}.}

\update{Both streams of research on GAI at work are nascent and have shortcomings that warrant further research. First, most existing research evaluates effects on categories of workers or professions, neglecting workers' in-depth, individualized practices \cite{Eloundou_2023}. Second, prior studies utilizing experimental methodologies do not reflect natural work settings, in which complex interdependencies with other tasks and roles are prevalent. Lastly, prior studies placed GAI at the center of their investigation and explored its implementation on workers' outcomes, rather than focusing on how workers derived their lived experiences from using these tools.}  

HCI research on GAI has largely focused on designing and developing systems leveraging GAI to aid workers' tasks \cite{zhou2024}. Other studies have used speculative design methods in controlled settings to gauge workers' preliminary perceptions and attitudes about tentative features to understand their potential benefits and/or harms in creative work processes \cite{wang24, wilcox24, kim2024}. In the context of writing, studies explore developing and implementing GAI to improve workers' outputs \cite{ Reza2024, Hoque2024}. For instance, \citet{lee22} developed CoAuthor, which assisted writers in drafting sentences and, in the process, helped them compose short stories. HCI studies also focus on augmenting workers' processes instead of directly affecting their outputs \cite{chung22, Benharrak2024, arakawa23}. For example, \citet{chung22} designed and evaluated a GAI-based system that enhanced writers' progress by offering context-aware cues and encouraging them to resume their work after interruptions. While workers recognized the advantages of these features, they also expressed concerns about the technical proficiency needed to effectively utilize custom-made GAI tools in their work. In response, several studies used off-the-shelf GAI technologies (e.g., ChatGPT) to understand the benefits and liabilities these tools brought to individuals' work \cite{Li2024,Noy_Zhang_2023, kadoma2024}. \citet{mirowski2023} conducted an experimental study with screenwriters that revealed tensions between the writers' creative instincts and GAI's functionalities, highlighting the limitations of decontextualized evaluations of such systems and underscoring the need for a deeper understanding of GAI integration in writing fields. 

Studies following the tradition of reflective HCI research~\cite{dourish2004} responded to this call by taking a more critical stance towards understanding creative workers' sociotechnical experiences in their natural settings \cite{Varanasi_25}. These studies focused on the early adoption experiences of creative professionals more broadly, such as game developers, trainees, and User Experience (UX) designers, about using off-the-shelf GAI technologies in their daily workflows~\cite{boucher2024,Suh2024,Kobiella2024}. Key findings indicate that the adoption of GAI is influenced by several considerations. One such consideration is the operational modality of these GAI (e.g., text-generation vs. image-generation) and the ease of working with the modalities in existing workflows \cite{boucher2024}. Some creative roles (e.g., programmers) found it more straightforward than others (e.g., graphic artists). For example, \citet{mim2024} showed how Bangladeshi graphic artists had to deviate from their traditional process of visualizing creative ideas on a canvas, instead putting significant effort into translating their visual concepts into textual prompts for GAI tools to generate visualizations. This shift in processes led to cognitive strain. However, research in this area remains limited within the context of writing professionals. An exception is \citet{Dergaa2023}'s meta-synthesis, which examined the writing practices of academics. Their findings suggest an overall increase in work productivity, albeit at the expense of authenticity and credibility. They emphasized the need for more in-depth research to understand how GAI elicits such trade-offs by capturing changes in work practices.

Taken together, prior research makes it difficult to discern whether the practices are driven by the novelty of GAI, or reflect routines that have been \textit{established} and \textit{sustained} as users become accustomed to them. Furthermore, several studies that captured lived experiences also involved young professionals whose workflows may not have matured enough to demonstrate complex strategies with GAI. Lastly, these studies disproportionately focus on adoption practices, limiting scholars' understanding of writing professionals who have actively \textit{resisted} GAI while continuing to shape their work in anticipation of its disruptive impact on the industry. Limited insights in this direction come from professionals engaged in knowledge and creative work \cite{Kobiella2024, boucher2024}, indicating that their resistance was primarily driven by skepticism about whether GAI would add value to their work. Research in this space could benefit from more in-depth treatment of resistance practices beyond professionals' initial speculations around their fears. In order to fill these gaps in research on GAI in the field of writing, we address the following two research questions: \\
\textbf{RQ-1:} \update{\textit{What resisting practices are writing professionals employing in response to GAI technologies in their field?}}  \\
\textbf{RQ-2:} \update{\textit{How do these practices compare and contrast with the embracing strategies of writing professionals?}}

\subsection{Job Crafting: A Theoretical Lens} \label{RW-jobcrafting}
To answer our research questions, we applied the theoretical framework of \textit{Job Crafting}. Job crafting originated as a theory of work design to understand the relationship between the design of people's work responsibilities, activities, and relationships and the way they derive identity and meaning from their work \cite{Wrzesniewski_2001}. What sets job crafting apart from other work design frameworks is that it attributes agency to workers, rather than managers or the organization, in designing their work. In this regard, the framework focuses on the bottom-up physical and cognitive changes that workers initiate to modify their role or relationship boundaries \cite{Tims_2012}. 

Wrzesniewski and Dutton introduced the idea of job crafting by showing how professionals invest significant energy in reconfiguring and shaping their jobs \cite{Wrzesniewski_2001}. Later research identified two main aspects of the job that individuals craft. The first aspect is the role, including the tasks it involves and the way people perceive the job. The second aspect comprises \textit{relational} changes that influence how people interact with their colleagues at work. Lately, researchers have consider workers' goals, differentiating \textit{approach} crafting, which involves actively creating opportunities to align one's work role with one's professional preferences, strengths, and goals by expanding the role, from \textit{avoidance} crafting, which involves the practices individuals employ to reduce negative aspects of work~\citet{Bruning_Campion_2018, zhang_2019}. 

Workers who engage in proactive crafting behaviors have been shown to benefit from increased job performance, satisfaction, and well-being \cite{Tims2013,Gordon2018}. However, not all workers engage in job crafting to the same extent. This variation is influenced by factors, such as rank \cite{Berg2010}, degree of autonomy \cite{Berg2010}, and the resources available to workers \cite{Petrou2017}. These findings suggest that job crafting may be especially relevant to creative workers with greater skill and autonomy. Moreover, while it is extensively studied to understand how full-time employees in traditional organizational settings establish goals and work-life balance \cite{Bruning2019, Kossek2016, Gravador2018} or manage relationships \cite{Bizzi_2017}), job crafting is also being applied to understand non-traditional work contexts, such as gig work \cite{Dominguez2016, Wong_Fieseler_2021}. 

Yet there is limited understanding of how AI technologies are shaping workers' crafting activities. Studying changing worker-led practices related to GAI through the lens of job crafting is beneficial for multiple reasons. The job crafting lens positions agency in the hands of the workers, the actual users of the GAI technologies, as opposed to the management that might be responsible for adopting and designing the rules around the technology. The framework also allows us to balance the techno-deterministic narrative surrounding GAI with a social constructivist understanding of how individuals integrate and use GAI to shape their work. Lastly, it offers a systematic approach to examining nuanced sociotechnical practices and lived experiences at a micro-level, complementing economy-level assessments of GAI's impact on work. \update{Moreover, creative workers' autonomy and dependence on craftsmanship positions them at the forefront of job crafting in relation to GAI.}

%% file: 3.Methods.tex
\section{Methods}
To address our research questions, we conducted an IRB-approved interview study with writing professionals who were familiar with GAI-based writing tools and actively engaged with them in different ways. Data collection spanned four months (Mar-Jun'24). In this section, we describe the recruitment procedures, the demographics of the participants, and the analysis procedure.

\subsection{Participant Recruitment and Demographics}
In order to recruit professionals from various writing professions, we employed a quota sampling strategy. We started by partnering with our university's in-house publication and marketing team. Over multiple meetings, we presented our diverse recruitment criteria, including the type of role, industry, size of the company, nature of tasks, seniority level, and gender of the professional. The first author, along with the publishing team, designed the recruitment materials and shared them using multiple avenues, including social media platforms like LinkedIn, Twitter and Facebook, dedicated mailing lists for writing professionals (e.g., Writers Guild of America), gig-work platforms like Upwork, and specific writing subreddits. To enable some comparability across the participants' narratives, we focused our recruitment efforts within the U.S.A. The recruitment flyer contained a link to the screening survey that captured participants' demographics, their primary work responsibilities related to writing, their overall attitudes towards GAI technology and its use in their routines, and their interest in participating in the interview study (see the supplementary material).  

From the individuals who indicated their interest in participating in the study, we shortlisted and conducted semi-structured interviews with 25 participants. We prioritized diversifying the attributes of the participants while shortlisting them. For example, we carefully considered participants with varied exposure and perspectives on the use of GAI tools in their work, including those inclined to use them, those who had stopped using them, and those not inclined to use them. We selected participants who were aware of GAI tools for at least a year.

\subsection{Participant Demographics}
Out of 25 practitioners, 10 identified themselves as men, 12 women and 3 as non-binary. Participants comprised professionals from diverse writing fields, including social media writers, grant \& proposal writers, ghostwriters, paralegals, copywriters, poets, staff editors, brand writers, technical writers, content strategists, UX writers, SEO and marketing writers, content creators, publishing company owners, authors, screenwriters, academic writers, fiction writers, book reviewers, journalists, and business/press release writers. Participation came from a wide variety of domains, including law, entertainment, business, sports, internet and social media, technology development, non-fiction, and creative fiction. We also interviewed six practitioners who worked in multiple capacities (e.g., ghostwriter and marketing writer), including three who owned and worked in their own publication company that hired writers to publish online content. Two participants had a minimum educational qualification of a diploma, 12 had a bachelors degree, 9 had a masters' degree and 2 had doctorates. All the practitioners first encountered the GAI tools at least one year prior to the date of interview (min.= 12 months, avg.= 15.84, max.= 27). More details are available in the Table \ref{tab:org-participants}.  

\input{7.table}

\subsection{Procedure}
We conducted semi-structured interviews remotely via video calls. Before the interview sessions, we contacted the participants to obtain their informed consent and set appropriate expectations. This involved familiarizing participants with the interview study procedure and clarifying our neutral stance towards GAI tools and our overall objective of developing effective safeguards for professionals experiencing disruptive technologies. We also explicitly stated the voluntary nature of the study. 

The interviews lasted between 65 minutes and 1.5 hours (avg.= 77 minutes) and were conducted in English. Our interview protocol focused on capturing how practitioners changed their work practices in response to the proliferation of GAI tools in their occupation. We began the interviews by capturing the practitioners' daily work practices (e.g., ``\textit{What does your typical work day looks like}?'') and high-level attitudes towards specific GAI technologies. (e.g., “\textit{It has been more than a year since [GAI Tool] was introduced. How do you feel about it}?”). We then captured insights on how practitioners responded to GAI tools and re-shaped their practices over the past year (e.g., ``\textit{Can you give me an example of how you changed your daily work tasks in response to [GAI tool]?}''). The third section focused on capturing changes in their work relationships (e.g., “\textit{How did you navigate interactions with colleagues and clients who had a different view of [GAI Tool]}?”) as well as their efforts surrounding skill development in the post-GAI period (e.g., \textit{``can you give me an example of how you see the [GAI tools] hindering or assisting your efforts in improving your skills?''}). We concluded the interview after capturing their overall perspectives on the immediate future of their work role (e.g.,``\textit{How do you perceive the impact of AI tools on the broader writing industry and professional standards?}''). 

During the interviews, participants were encouraged to share their screen and show us specific examples from their work tasks. The interviews were recorded with the participants' permission. After every few rounds of interviews, the authors revised the interview protocol collaboratively to capture deeper insights relating to our research questions. We provided compensation of \$50 in the form of an Amazon gift voucher or an equivalent contract (e.g., on Upwork) for study participation. We stopped our interviews once we reached theoretical saturation in our data. 

\subsection{Data Collection and Analysis}
Overall, we collected 32.5 hours of audio-recorded interviews (transcribed verbatim). We also captured several pages of notes during the interviews, especially for three participants who denied permission to record the interview. All of this data was analyzed using inductive thematic analysis \cite{Braun_Clarke_2006}. As a first step in our qualitative analysis, we started taking multiple passes of our transcribed data to understand the breadth of the interviewee's accounts. Subsequently, we conducted open-coding while avoiding any preconceived notions, presupposed codes, or theoretical assumptions. During this stage, we also captured our reflections in the form of memos. The process resulted in 82 codes. Through collaborative discussions, we removed overlapping codes and discarded the duplicate ones. The resulting codebook consisted of 61 codes. Examples included \textit{task reconfigurations}, \textit{increased efficiency}, \textit{expanding role}, and \textit{increased collaboration w/ AI}. As a second step, we used theoretical coding \cite{Hennink_2019} to develop a sense of thematic organization of our codes. Using an abductive process \cite{Timmermans_2012}, we reflected on the key elements of job crafting theory \cite{Lazazzara_2020, Wrzesniewski_2001}, namely role and relational crafting as well as avoidance and approach crafting, to further map, categorize, and structure the codes under appropriate themes. To establish validity and to reduce bias in our coding process, all the authors were involved in prolonged engagement over multiple weeks. Important disagreements were resolved through peer-debriefing \cite{Creswell_Miller_2000}. Examples of the resultant themes included, \textit{factors impacting GAI adoption}, \textit{task-re-planning around GAI work}, \textit{reducing collaboration with network}. The final codebook is provided in Appendix \ref{app:codebook}. Based on these final themes, we present our findings in the next section.

\subsection{Positionality}
In this study, our objective was to engage with writing professionals holding diverse perspectives, including those who embraced and those who resisted AI, to capture a comprehensive account of their work transformations. To maintain neutrality, we ensured that our communication remained unbiased, neither favoring nor opposing GAI, and that participants fully understood our research while proactively addressing any concerns they had. We also recognize the importance of studying and uncovering these worker-led practices as a critical step toward ensuring equitable outcomes. To achieve this, we drew on our diverse research experience in Human-Computer Interaction (HCI), Organizational and Management Studies, and Computer-Supported Collaborative Work (CSCW) for our analysis and interpretation. Lastly, two authors identify as men, and one as a woman. Two authors hold senior faculty positions at a well-established university in the Global North, while the third is an early-career scholar at the same institution.

%% file: 7.table.tex
\begin{table}
 \center
 \renewcommand\arraystretch{1.3}
 \footnotesize
 \begin{tabular}[t]{|p{.55in}|p{2.43in}|}
 \hline
 \multicolumn{2}{|l|}{{\bf Total Participants} (n=25)}\\ 
 \hline
  Gender & 
       \begin{tabular}{lll}
           Women:12  & Men: 10 & Non-binary: 3 \\
       \end{tabular}\\
\hline
 Age (years) & 
       \begin{tabular}{llll}
         Min: 24 & Max: 65 & Avg: 42.6 & S.D: 12.6\\
       \end{tabular}\\
\hline
Education (degree)  & 
       \begin{tabular}{llll}
           High school: 2 & Bachelor's: 12  &  Master's: 9 & Doctoral: 2\\
       \end{tabular}\\
\hline

Job type & 
       \begin{tabular}{lll}
           On payroll: 12; & contract: 8 & freelance: 7  \\
       \end{tabular}\\
\hline
Region  & 
       \begin{tabular}{p{.4in}p{.5in}p{.4in}p{.4in}}
           Eastern USA: 6 & Southern USA: 4 & Western USA: 9 & Central USA: 6\\
       \end{tabular}\\
\hline
Experience (years) & 
       \begin{tabular}{llll}
           Min: 4  & Max: 35  & Avg: 17.8  & S.D: 10.6 \\
       \end{tabular}\\

\hline
Roles & 
       \begin{tabular}{p{2.3in}}
          Journalist (x2); Paralegal (x2); SEO editor (x2) ; Social media writer (x2); Grant Writer; Ghostwriter (x2); Owner (x2); Marketing writer; Staff editor; Poet; Production Editor; Technical writer; Copywriter; Author (x2); Content Strategist; Fiction writer; Book Reviewer; Brand Writer; Screenwriter; Business writer;  \\
       \end{tabular}\\
\hline
GAI exposure (months) & 
     \begin{tabular}{llll}
           Min.: 12 & Max.: 27 & Avg.: 15.84 & S.D.: 12 \\
       \end{tabular}\\
\hline
GAI tools & 
       \begin{tabular}{p{2.3in}}
          ChatGPT (x20); CoPilot (x5); Writesonic (x1);, Elicit (x2); CoCounsel (x2); Anyword (x3); Jasper (x1); Claude (x2); Custom-GAI (x3); Neuron (x2); AlliAI (x1); Waldo (x1) \\
       \end{tabular}\\

\hline
\end{tabular}
\caption{Demographic details of writing professionals interviewed for the study.}
 \label{tab:org-participants}

\end{table}

%% file: 4.1.expansion.tex
\section{Findings}

\begin{figure*}
    \centering
    \includegraphics[width=.7\linewidth]{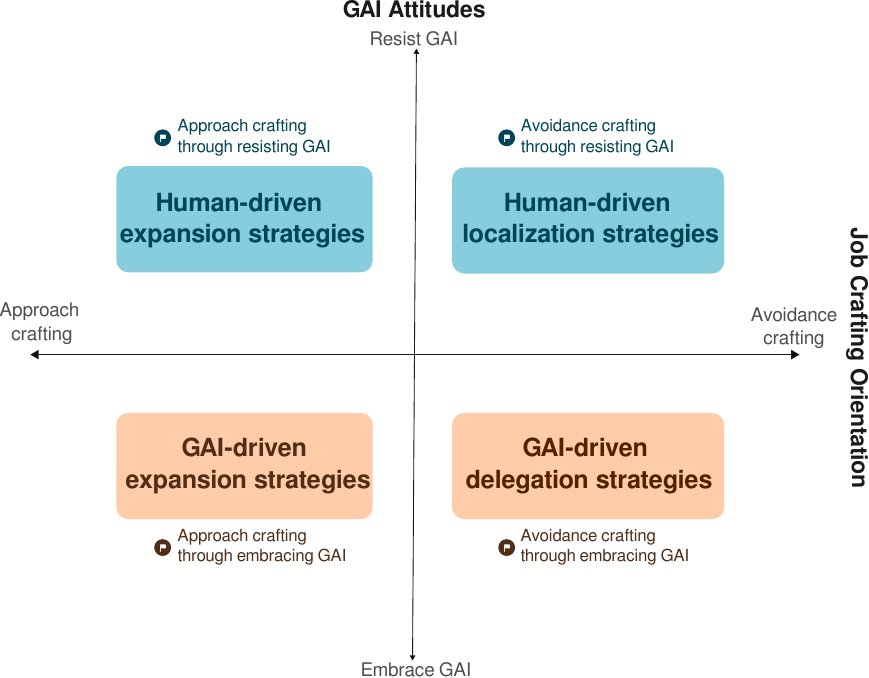}
    \caption{The image illustrates four types of strategies observed among different writing professions. The x-axis represents the crafting orientation derived from Job Crafting theory \cite{Wrzesniewski_2001}, differentiating approach crafting from avoidance crafting. The y-axis represents the attitude towards GAI, differentiating responses related to embracing GAI from responses resisting it.}
    \label{fig:quad-image}
\end{figure*}

\up{As expected based on prior research \cite{vimpari23, boucher2024}, writing professionals we studied identified threats and opportunities associated with GAI, albeit to varying extents. Extending prior work, our findings reveal two unique forms of bottom-up, preemptive strategies professionals employ to proactively reshape and adapt their roles in anticipation of these changes (see Figure \ref{fig:quad-image}).}

\up{The first set of strategies focused on \textit{role expansion} through approach crafting emphasizing increasing the desirable and motivating aspects of their roles \cite{Bruning_Campion_2018}.  Professionals with higher perceptions of threat, who resisted GAI, employed \highb{human-driven expansion strategies} to enhance their human-centric tasks and skillsets, with the explicit goals of underscoring the distinct value of human labor and differentiating themselves from GAI users (see top-left quadrant in Figure \ref{fig:quad-image}). Conversely, professionals more inclined to embrace GAI employed \higho{GAI-driven expansion strategies} to leverage GAI in enriching their work and introducing innovative elements to their roles (see bottom-left quadrant in Figure \ref{fig:quad-image}). We detail and compare these role expansion strategies in Sections \ref{expansion-1} and \ref{expansion-2}.}

\up{The second set of strategies involved \textit{streamlining} one’s role through avoidance crafting \cite{Bruning_Campion_2018}. These strategies sought to reduce uncomfortable or undesirable aspects of their roles while safeguarding core responsibilities. Professionals who resisted GAI employed \highb{human-driven localization strategies} to minimize unnecessary human effort, ensuring competitiveness against GAI users (see top-right quadrant in Figure \ref{fig:quad-image}). On the other hand, those who embraced GAI adopted \higho{GAI-driven delegation strategies} to streamline and delegate mundane or repetitive tasks to GAI (see bottom-right quadrant in Figure \ref{fig:quad-image}). We elaborate on and compare these role streamlining strategies in Sections \ref{streamline-1} and \ref{streamline-2}.}

\subsection{Expanding Role Through \texorpdfstring{\highb{Human-Driven Expansion Strategies}}{Human-Driven Expansion Strategies}: Strengthening Identity, Skillsets, and Practices} \label{expansion-1}

\subsubsection{\update{Strengthening Professional Identity \& Ownership}} \label{identity}
\begin{figure*}
    \centering
    \includegraphics[width=1\linewidth]{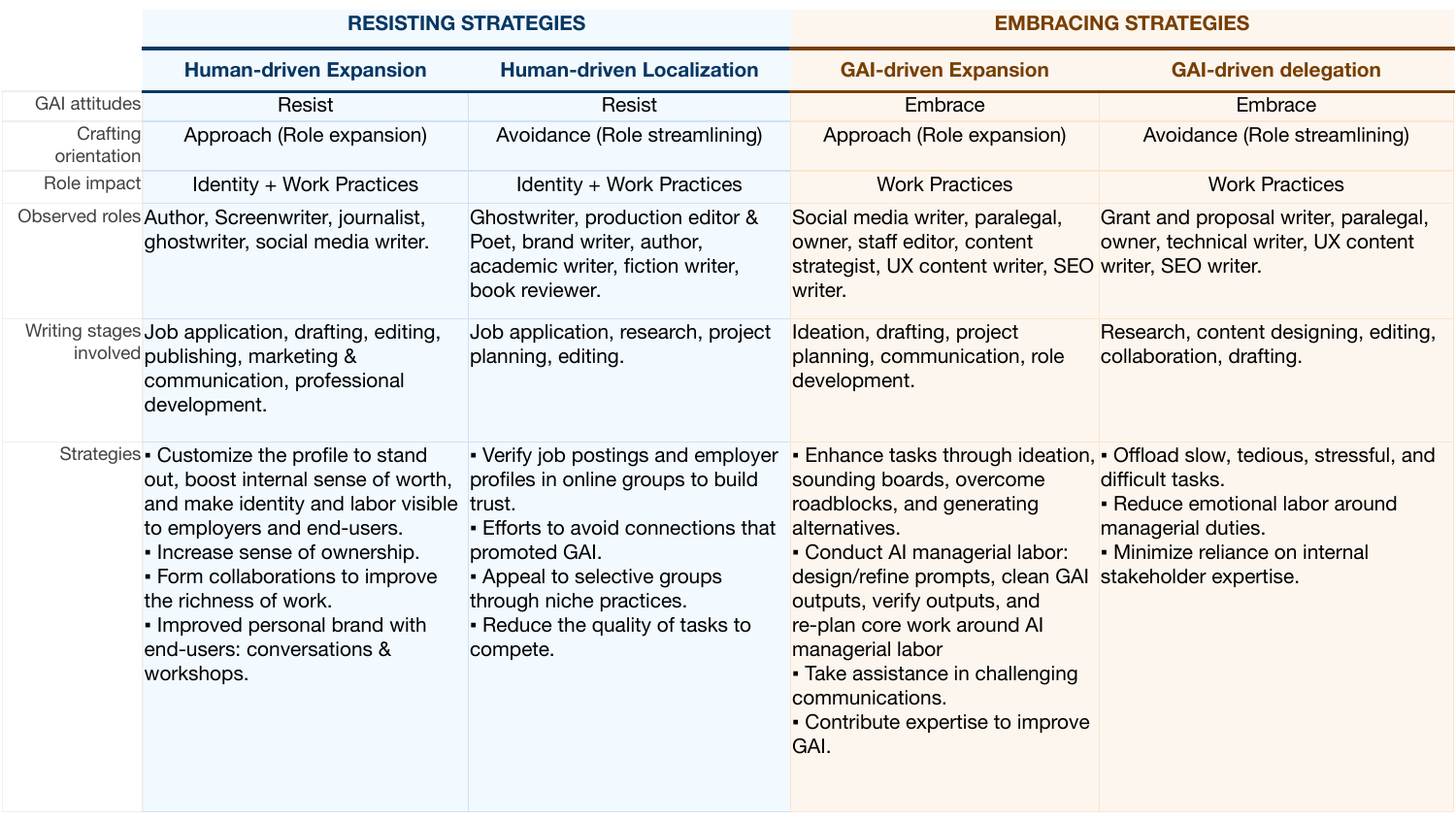}
    \caption{Comparison of four strategies used by writing professionals in response to GAI. The left two columns represent resisting strategies, while the right two columns represent embracing strategies.}
    \label{fig:strategiestable}
\end{figure*}

\update{Professionals employing \textit{human-driven expansion strategies} used crafting techniques to emphasize and enhance their \textit{worker identity} and professional worth at various stages of their work, responding to the growing influence of GAI in their fields (see Figure \ref{fig:strategiestable}).} Beginning with their applications for writing jobs, these professionals invested significant effort in tailoring their profiles and interacting with employers to distinguish themselves from those relying on GAI-generated content and to exhibit their skill to impress potential clients or employers. P25, a full-time and freelance SEO writer shared: 

\begin{quote}
    \textit{\update{``I take time and include personalized cover letters with all my bids. The trick is to make small anecdotal details matching the clients' proposal. \dots sometimes a few pieces from my past projects \dots Good humor in these letters goes a long way to stand apart! \dots I even changed my profile picture with one where I am teaching my daughter to write \dots they [clients] should not feel I am an AI bot.''}}
\end{quote}

\up{We observed these activities to be more prevalent in job roles that engaged in part-time or freelance work, such as social media writers and script writers, among others.} Professionals using these strategies exhibited similar practices while engaging in their work tasks, \update{actively finding ways to make their human identity explicitly visible in their labor to their employers}. This was especially evident in cases where employers introduced GAI technologies (officially or unofficially) into the workstreams of the professionals we interviewed (n = 6). In response, P20, while working on articles as part of his journalistic responsibilities, maintained a detailed informal log that he shared with his editor, clearly distinguishing between the articles he wrote himself and those in which he used GAI technology. When asked why he kept this record, \update{P20 explained}, 

\begin{quote}
    \textit{``\update{The [GAI] tool is not made by my organization.\dots My editor does not have a backend access to my account to see my prompts and know whether I wrote the press release or the [GAI] tool spit it out \dots Did I do the research, scour the backstories, or did interviews for the byline? \dots I prefer to make sure my editor knows 100\%  that it is my work \dots not just the AI's \dots and not question my work ethic.} ''}
\end{quote}

\up {
%Additionally, P20 believed that maintaining such a log would help his manager recognize the value of the effort he invested in his work.
Additionally, within these logs, he included only the \textit{``core''} writing tasks, those traditionally valued in journalism, such as investigative pieces and feature articles, while excluding press releases and routine event coverage.}

Surprisingly, even when their contributions were not visible to their managers, some professionals still insisted on applying their own efforts to tasks to elevate their \textit{sense of worth}, despite the additional time required to deliver the results. P10 worked as a brand writer for a marketing agency in the beauty and fashion industry. When her organization abruptly shifted to using GAI to scale content production, they required writers to first generate content with GAI and then edit it to increase the pace. P10 found this approach demeaning to her human labor. She shared:
\begin{quote}
    \textit{``Sometimes, just to feel better, I sneak in a couple of my own paragraphs, even if it takes more time, because it keeps my writer spirit alive.''}
\end{quote}

% These nuanced behaviors of P10 and P20 show that professionals were comfortable employing resistance strategies for certain core tasks while simultaneously adopting embracing practices for more mundane ones.  
These nuanced behaviors of P10 and P20 challenge prior research on resistance strategies, which often frames them as clear-cut and resolute rather than a delicate balancing act \cite{Jochum_2023}.

\update{Professionals also invested in ways to better retain control over their authored work, protecting both themselves and their creations from the influence of GAI.} These practices included finding ways to generate more content for which they could retain \textit{ownership} while also establishing long-term revenue streams for such articles. P25 shared how they sought ways to earn money as a writer in new ways: 

\begin{quote}
    \textit{``I'm taking steps to be more independent \dots Writing itself isn't going anywhere \dots This year, me and a few of my peers have started focusing on how we can own the content. If everything hits the fan, I still want to be able to write but I'm going to own that. I have started 11 websites. The baking website is my best contender; it has Facebook followers. I am working on creating books on Amazon KDP (Kindle Direct Publishing), and send them on mailing lists to be able to sell cookbooks and courses to the followers. ''}
\end{quote}

% Moved from relationship 
\update{Professionals also focused on making their identity and work process more visible to their readers}. To achieve this, they initiated multiple communication channels, both online, through social media platforms like Instagram and TikTok, and offline, at various meetups. For instance, P16, an author of two best-selling non-fiction books, regularly attended book readings and literary meetups to share anecdotes about his writing process. Similarly, P6, an immigration paralegal, started offering free introductory tutorials for clients, demonstrating the complexity and dedication involved in the preparation of immigration petitions. P23, a screenwriter for online long- and short-form content, shared:

\begin{quote}
    \textit{``Having a relationship with the community that you serve, with your sources, making yourself sort of indispensable \dots it does sort of tell the management that if we get rid of this person, we're losing somebody who is trusted in the community \dots people are consuming stories just because of who we are and why we produced it, less about what the topic is.''}
\end{quote}

\up{Overall, these identity-focused strategies extend the notion of \textit{identity work} \cite{Dunn2024}, an established practice among creative professionals, by demonstrating how the proliferation of GAI has significantly intensified its use as a resistance strategy. One notable indication of this shift is the growing reliance on personal narratives (e.g., informal logs used by P20) beyond traditional work outputs, a trend also observed during the post-platformization era of many creative roles~\cite{munoz22}. }

\subsubsection{\update{Expanding Skillsets \& Practices}} \label{skillsets-1}
\update{Professionals engaging in human-driven expansion also invested efforts in expanding their work practices, alongside their identities, to safeguard themselves against the influence of GAI. They shifted their focus from solely meeting their employers' expectations to prioritizing the value they provide to readers consuming their content. To achieve this, they leveraged uniquely human capabilities that GAI lacks to demonstrate a distinct competitive edge over GAI-generated content.} A common tactic was to counter GAI's inability to generate rich, localized content in their written pieces for readers to consume. P21, another journalist working for a local newspaper, shared how he and his team implemented these strategies in real-time. To combat the competition from publication outlets relying on GAI-based content farms, P21 took initivates to create hyperlocal versions of key articles, incorporating local information such as up-to-date commentary, details of local businesses relevant to the topic, and interview excerpts from local citizens. He felt that such localization efforts were also helpful in achieving better indexing on Google searches.

\update{Another effort was to rely on the strength of their human connections to counteract GAI technology's ability to generate content rapidly.} Through collaborating with those in their networks, professionals enhanced the depth of their work by incorporating human experiences and insights. They perceived that such approaches improved the overall richness and relevance of their deliverables. For this purpose, they organized collaboration initiatives, such as workshops and panels in their organizations and peer networks. For instance, P20, in his organization, pushed for ``double by-line stories'' \footnote{news articles written jointly by two or more authors.} that gave him the opportunity to produce collaborative content in his organization that included topics and insights that he could not think of on his own. He felt that GAI models tend to exhibit a ``filter effect,'' producing skewed or biased information, which human collaborations can help circumvent:   

\begin{quote}
    \textit{``%\dots When I was working with the other folks, we were seeing issues with each others' perspectives brought out by our complementary expertise and training \dots 
    We wrote a piece on effects of air condition on the environment where my coworker [who specialized in consumer reports] wrote about how people were using them. I wrote the climate aspect of it, and we worked with an economist who wrote the cost-benefit aspect of it. We were combining all our perspectives from different fields. That synergy was really useful in making connections that weren't as apparent \dots [and] drawing conclusions that were well-rounded \dots We are pushing hard for such real human collaborations.''}
\end{quote}

These practices reflect collaborative crafting \cite{grant2017re}, a specialized form of job crafting in which individuals come together to collectively shape their work. \up{However, what makes this form of collaborative crafting unique in our context is the emphasis on preserving professional identity (e.g., environmental expert) and the specialized knowledge that individual professionals contribute to the collaboration, alongside the traditional goal of generating collaborative insights through this form of crafting}. 

To further enrich their deliverables, some professionals took it to the next level and learned complementary skills from neighboring roles, such as graphic design for journalists, UX design for UX writers, and content design for grant writers and authors. P16 shared his views:

\begin{quote}
    \textit{``If I'm the person who writes the stories, shoots the video of the oral history interviews, and takes photos, all of a sudden you're doing too many tasks that involve you wearing too many different hats to be easily replaceable with these [GAI] tools.''}
\end{quote}

Developing broader skillsets gave them confidence, increased their self-esteem, and reduced the perceived risk of job loss. \up{We observed these practices taking shape in both traditional writing roles as well as more contemporary roles.}

\subsection{Expanding Role Through \texorpdfstring{\higho{GAI-Driven Expansion Strategies:}}{GAI-Driven Expansion Strategies:} Strengthening Practices}
\label{expansion-2}

\subsubsection{\update{Enhancing Creative Workflows}} \label{expansion-2-enhancing}
\update{In contrast to human-driven expansion strategies, professionals who incorporated GAI-driven expansion strategies mainly focused on enhancing their work \textit{practices} to improve the overall efficiency and quality of their work (see Figure \ref{fig:strategiestable}).} One such practice was using GAI to generate ideas and ``jump-start'' creative ideation and writing processes. 
%For example, User Experience (UX) writers in our study routinely used ChatGPT  the content that is integrated into the user interfaces of digital technologies.
% This process requires UX writers to develop persuasive language while navigating the unique constraints of newly designed interfaces. P13, a UX writer in a large corporation, shared: 
% \begin{quote}
%     \textit{``Word placement in designs is not like traditional writing \dots it is extremely tricky and I struggle a lot with finding the right words to say. It is much easier if you can talk about it in a longer document because you can explain the rationale. It's harder when the words you're trying to say are three words in a UI with a free space of only 100 pixels. So I have to think a lot to figure out those words. With ChatGPT, I can [iterate] by just asking for options \dots it helps jump-start as I often don't have to spend so much time ideating. \dots The faster I do that and get feedback, the closer I get to the deliverable.''}
% \end{quote}
% Incorporating ChatGPT's input allowed P13 to boost her workflow. 
Even in situations where professionals did not end up using the ideas generated by GAI, those ideas still served as useful starting points or sounding boards in the subsequent steps of their writing process. For instance, screenwriters often create a beat sheet, outlining key beats, as part of their story development. When P1, a social media writer who did part-time script writing, encountered writer's block, he used ChatGPT to generate the initial beats for his scripts. Although he didn’t incorporate the exact beats generated by ChatGPT, using them as a reference made it easier for him to develop his own subsequent versions. 
%Pontetial replacement of text with quote (optional)

Professionals also used GAI to improve the quality of their overall tasks by exploring GAI-recommended alternatives in the early stages of their work. \update{For instance, P1 shared how he used GAI to generate multiple writeup ideas for the descriptions he used in search engine optimization}:

\begin{quote}
    \textit{``I actually go into the client's WordPress site, editing the metadata. ChatGPT helps me generate super good meta descriptions...I just type, ``come up with 10 meta descriptions’’.  I then piece together what I like from those, and then use the ones I love \dots it's been amazing for its speed in coming up with alternative seed keywords. It's really useful when exploring multiple complex domains as part of the keyword strategy.} 
\end{quote}  

Professionals also used similar conversations with GAI to overcome obstacles and make progress in their workflows, such as writer's block arising in the middle of a creative project. P3 is a ghostwriter who worked in a publishing company that produced content for children. \update{While P3 resisted using GAI in his creative work, he explained that when encountering creative blocks, he occasionally used ChatGPT to brainstorm ideas, yielding variable results}. P3 shared one such instance: 
\begin{quote}
    \textit{\update{``I was writing this story about this shy duckling that was going to the school for the first time \dots the theme of the story was around learning disability \dots I wanted to name the school [something] that fit this whole narrative and the character arc while being a little funny. It's a very small thing but I was sitting on it for hours! \dots I finally gave up \dots used ChatGPT for options. None of them were particularly great but it did make me think in a few new directions.''}}
\end{quote}

Although P3 did not end up using any of the options suggested by ChatGPT, the alternatives gave him an idea of his own that he eventually implemented. These findings reveal that even professionals resistant to GAI occasionally adopted its strategies when faced with creative limitations, highlighting how resistance and adoption can coexist in writing. This offers nuanced insights into how other creative practitioners, such as art hobbyists and game designers, may integrate different GAI strategies rather than using them in isolation~\cite{sanchez23, vimpari23}.

% from relationship

Professionals also used GAI to develop, sustain, and improve their professional relationships by drafting communications to other people. Several professionals (n= 5) reported using GAI to help write emails, chats, and posts when dealing with challenging interpersonal situations at work. For instance, P14, a freelance copywriter working in the health domain, recalled using ChatGPT to explore alternative ways of expressing his negative emotions with a client while avoiding harm to the relationship:

\begin{quote}
    \textit{``\update{When you are in my domain [freelance copywriting], there is a lot of back and forth with the clients in the writing phase, especially during lead generation or finalizing the USP (unique selling proposition). This client changed their request at the last moment and I lost my cool \dots [before this] I used to compose these messages in that [angry] emotional state and that exacerbated the issue. \dots [Now] [With GPT] I dump my thoughts and ask it to generate a message with keywords like professional, empathetic, solution-oriented. \dots I have created a separate GPT that just does that. ''}}
\end{quote}

Professionals adopting GAI-driven expansion strategies also advocated for expanding their professional boundaries in addition to shaping their own roles. Three professionals we interviewed enacted this expansion by using their expertise to evaluate and enhance the outputs of GAI models in the context of paid or unpaid programs, in addition to completing their regular writing tasks. In so doing, they applied their field-specific expertise as seasoned professionals to broadening the capabilities of GAI, and then shared the learnings from these experiences with professional colleagues, strengthening human networks. \update{P4, who collaborated with a pioneering company in the GAI space, developed new GAI expertise in this way and regularly shared the insights with her professional networks. P4 described},

\begin{quote}
    \textit{\update{``I enrolled in this expert evaluation project for [a big tech company] their multi-layered prompts. My job was to figure out at which level the [G]AI started giving inaccurate information and then correct it. \dots Nothing fancy and it does not pay me anything, but I get to be at the forefront of the innovation. \dots I am part of these two active paralegal associations where I share these experiences and help my colleagues also learn about these innovations. ''}}
\end{quote}

\up{Interestingly, P4 shared that several members of these paralegal associations who actively read her experiences preferred not to use GAI, which did not deter her from sharing this information. This evidence shows inherent dependency between professionals who resist and those who embrace GAI within the same profession that are often invisible.}

%%%% From AI Managerial labor
\subsubsection{Incorporating AI Managerial Labor}
Professionals who used GAI-driven expansion strategies also took initiative to integrate and optimize GAI use within their workflows to achieve effective outputs. We group these efforts under \textit{AI-managerial labor}. As part of this effort, professionals engaged in prompt engineering, a well-established practice of designing and refining the input prompts to receive beneficial outputs \cite{Varanasi_25, openai_guide, mckinsey_eng}. 

This required substantial effort and time from professionals, demands they managed by re-organizing their entire workflow to keep AI-managerial tasks together for a particular project. For example, P6, a paralegal in a large firm specializing in immigration issues, was drafting a client's green card application. As part of the process, she had to develop a business plan for the client. Having limited awareness of how to do this, she used ChatGPT to conduct market analysis and generate financial projections, which were critical to the business plan. To optimize ChatGPT within her workflow, she found it necessary to consolidate the AI-managerial tasks into a single session. She shared: 

\begin{quote}
\textit{\update{``With ChatGPT, I need to block two to three hours to complete the prompting that I started, or at least the majority in one sitting. If I leave it mid-way, once I am done with a few prompts, and come back to it after a few days, it is extremely difficult to follow the reasoning as to why I provided the particular follow-up prompts \dots if I re-run or continue working from that point, there is no guarantee that ChatGPT would provide similar output in the same line of logic again!''}}
\end{quote}

\up{These findings illustrate how certain research-heavy technical writing professions can develop higher dependency on GAI tools, pushing them to reshape their work in major ways.}

%% file: 4.2.delegation.tex
\subsection{Streamlining Role through \texorpdfstring{\highb{Human-Driven Localization Strategies:}}{Human-Driven Localization Strategies:} Creating Niche Identity \& Reducing Effort}\label{streamline-1}
\update{Similar to human-driven expansion strategies, professionals following human-driven localization strategies also focused on shaping both their identity and their practices by concentrating and streamlining their efforts (see Figure \ref{fig:strategiestable}).}

\subsubsection{\update{Creating Niche Identity}} \label{niche}
To protect the most treasured aspects of their professional identity, professionals employed human-driven localization practices, including avoiding GAI-related jobs and distancing themselves from individuals who advocated for GAI use, to preserve their credibility within their community. For instance, participants restructured their professional relationships by changing the way they searched job descriptions to avoid those likely to involve GAI. A primary indicator was the keywords sprinkled across the job descriptions, such as ``need to use AI'' or ``open to using AI''. Another indicator was employers' prior treatment of professionals' work or reducing their pay. Practitioners engaged in freelancing work gathered this information from reviews of employers on the platform and through their professional networks. Others relied on emotional, informational, and instrumental social support from close-knit peer communities on social media, such as Facebook (now Meta), to vet clients or employers before committing to work. When asked why they put so much effort behind vetting a particular client, P10 replied:
\begin{quote}
    ``\textit{In this day and age, it absolutely matters who I am associating with and securing work [from]. I do not want to be that person who is seen as pro-[G]AI. That is a sure-shot way to end your career before it starts.}'' 
\end{quote}
One such group that P10 interacted with maintained a curated list of clients, companies, and organizations that aligned with the community's core values and were considered ``safe'' to work with.  

Professionals also proactively \update{reframed their role around craftsmanship, and used their craftsmanship capabilities to appeal to a more selective subset of groups (employers, clients, or communities) that valued human labor and its resultant artifacts}. These groups enforced stringent standards, such as using AI detectors (e.g., GPTZero \footnote{https://gptzero.me} or ZeroGPT \footnote{https://www.zerogpt.com/}) to examine the source of the written content. Professionals distinguished their work with keywords such as ``boutique'', ``niche'', or ``organic'' to appeal to these selective employers or professional communities and create a niche for their craft. For example, P7, an author and poet working for a children's publication company, articulated parallels between this change in practice to position her work as more boutique and the evolution of market dynamics in the shift from printed books to e-books. She shared, 
\begin{quote}
    \textit{``Everyone thought that print books will die because of e-books. That is not the case \dots books have only increased their value as tangible objects because certain publishers chose to make it a boutique object, among other things. \dots \update{It boils down to how can 100\% human generated content, like poetry distinguish itself? \dots I am seeing a shift in publications within my community where they are clearly aligning themselves to these core values of increasing the value of human-generated content with rules like no AI written content, etc.}''}
\end{quote}

\up{These findings echo the \textit{artisan approach}, a form of thinking developed in making, tinkering, and design fields \cite{zoran2014, Kera_2017}. Professionals following this approach tend to cultivate a deeper commitment to their craft through years of skill development, which motivates them to strongly distinguish their manual labor from commercial artifacts produced at scale.}

\subsubsection{\update{Reduced Effort in Work}} \label{reduced-effort}

Professionals who employed strategies within this category moderated their level of human effort without succumbing to the pressure of using GAI technologies. One common practice was to meet market expectations on the turn-around time for the required tasks by skipping a few steps in their work processes and reducing the quality of the deliverables (task crafting). For example, P17 shared how she modified her workflow to accommodate the changing market demands:

\begin{quote}
\textit{``I have no option but to say yes, because I gotta keep working. In an ideal world, I will not make any compromises. I have also written a script for someone and I got paid \$10,000 \dots Recently, the kind of work I am getting is in no way the same. Everyone wants it fast, quick, and offers far less money \dots I have just reduced the depth in research and the amount of iterations that I do with my work. They need something fast and cheap? I just give them that. Because if I don’t do it, they will go with the person who is using AI.''}
\end{quote}

\up{A possible downstream consequence of this strategy is that by reducing the quality of her human work, P17's employers may find it harder to differentiate human work from GAI work. Such responses can create a `race to the bottom,' intensifying pressure on professionals to rely on GAI. This can lead to the \textit{AI-ghostwriting paradox} \cite{draxler24} in which professionals, having used GAI, no longer view themselves as the sole authors of their content but remain hesitant to attribute authorship to GAI.}

\subsection{Streamlining Role through \texorpdfstring{\higho{GAI-Driven Delegation Strategies:}} {GAI-Driven Delegation Strategies:} Reducing Task Load and Dependencies}
\label{streamline-2}

\subsubsection{\up{Delegating Repetitive And Tedious Tasks}} \label{delegate}
GAI-delegation strategies were used by professionals to save time and improve the overall efficiency of their key workflow tasks (see Figure \ref{fig:strategiestable}). One way they did this was by offloading repetitive tasks. For example, several professionals working in social media, marketing, and search engine optimization (SEO) delegated information foraging tasks to GAI to reduce the amount of time \update{they were spending on their pre-writing activities, such as backround research}. These professionals engaged with a wide variety of topics in their domain. To develop an understanding of the topic they were writing about, they had to visit different hubs of information, such as news websites and databases (e.g., legal, medical), and read scholarly manuscripts. \update{P22, shared:}

\begin{quote}
    \textit{``\update{What I would do before [G]AI is visit all these different kinds of research hubs that were available. Whether it was a medical interest or business-related databases that I had access to, through libraries that I was a part of. Then I came up with my own terms to investigate if there were any sources that were relevant to this kind of thing. It was painstaking when you're just trying to get a feel for the landscape of a market. I have completely delegated this hours worth of work to [G]AI \dots it takes me several minutes and best. I don't have to waste money on the subscriptions.}''}
\end{quote}

For many of these professionals, several specialized GAI tools designed for information foraging, such as Writesonic, Elicit, and CoCounsel, were game changing, as the time they saved searching allowed them to focus on sense-making tasks.
%\up{Interestingly, this inclination to proactively reduce time spent on preparatory tasks, such as information foraging or preliminary exploration, is also evident in creative fields that are incorporating image-based GAI to support visual modalities, including graphic design \cite{palani24} and game development \cite{boucher2024}.} 
%The multi-modal capabilities of some GAI technologies also enabled professionals to delegate repetitive tasks like content layout and formatting to GAI. For example, P22, a grant writer and desktop publisher, shared how she was responsible for the overall design of grant proposals in her organization. She sourced information from multiple subject matter experts, some of whom provided materials in image formats, to draft grant proposals. She regularly used Microsoft's Copilot to reduce the manual task of converting content from images to text, avoiding the need to retype everything. 
% A few professionals also delegated tedious tasks that induced stress in their overall work (n=3). One particularly tiring task was copy-editing low-quality writing from external and internal stakeholders, which they found much easier to delegate to GAI technologies, thereby easing their burden. For example, P5, the owner of a publication micro-enterprise, introduced ChatGPT into his workflows to reduce the overhead of editing "poorly written stuff" produced by his junior writers, which had recurring issues.

\subsubsection{\up{Reducing professional dependencies}}
Professionals in this category also reduced their overall workload by eliminating certain professional dependencies, either partially or entirely. This practice was particularly prevalent among senior professionals (n=4), such as copy editors and owners of small-scale publication houses, who managed other writers and oversaw project deliveries. Our findings indicate that these professionals utilized GAI to alleviate the emotional labor involved in managing workers in writing roles and the additional burden of correcting work. For example, P15, the owner of a publishing house producing science-based content, described how they replaced their entire writing staff with ChatGPT. This effort streamlined their operations and reduced the effort required to manage writers by automating the staff's work: 

\begin{quote}
    \textit{``The difference now is that I'm not dealing with a lot of writers; I'm not giving them therapy sessions when they are falling behind, [or] if I rejected their manuscripts \dots With ChatGPT, it's easy to keep beating on it and make it change. I couldn't do that with my writers without striking a nerve \dots I had them at very hard pace \dots I explained every edit, they spent two days doing that, and still missed the deadlines. I ended up using ChatGPT to do all the things that they were doing. My effort has reduced.''}
\end{quote}

Participants also reduced professional dependencies on internal stakeholders to shorten the overall turnaround times. They shifted from relying on experts at every step of the process, who were often busy and slow to provide insights, to using GAI for early iterations of their work. P13 shared how she used GAI to reduce her dependence on subject matter experts by asking ChatGPT some of the questions for the early draft of her UX reports (e.g., ``\textit{What are the permits required to start a hotel business in New Jersey?''}), which would otherwise have required expert opinions. P13 shared:

 \begin{quote}
     \textit{``\update{I was working with a government agency on their new app on zoning regulations \dots It takes 4-5 days to get a response from an SME (subject matter experts) who knows about zoning. Without that knowledge, I cannot write key parts of my report \dots  I used ChatGPT and generated the first version of the draft that I submitted to the SME \dots I would ask it questions about certain things like, `imagine I am a CEO of hotel chain, how do I start a hotel business in New Jersey?' It usually spits out useful information that I can then build on and integrate in the report and get some feedback.''}}
 \end{quote}

\subsubsection{\up{Offlading Difficult Tasks}} \label{imp-tasks} A more concerning trend we observed was that several professionals (n=4) delegated tasks they found difficult and had little interest in developing skills for, despite these tasks being important to their professional duties. One such task that was common across different roles was the ability to incorporate diverse ``voices'' into their writing styles. This process includes embedding specific jargon, diction, or tone to which readers could relate. For P5, one of the biggest challenges in his marketing campaigns was capturing the voice of his target audience and increasing article engagement through the right choice of words and style. In one particular case, P5's client needed targeted content that had to be ``\textit{bubbly}'', ``\textit{techy}'', ``\textit{energetic}'', ``\textit{the kind that gives vibes that we wake up at 3 AM, meditate, and are extremely productive}''. Although P5 could easily recognize this type of content, it was challenging for him to compose it from scratch as he was trained as a technical writer. P23 had a similar challenge when she was writing scripts for short-form online videos aimed at young adolescents. P23 shared: 
\begin{quote}
    ``\textit{For developing this character's voice meant sitting and watching multiple existing shows and documentaries around the topic \dots I passed an initial draft with the voice requirement in the [ChatGPT] tool to capture some of these voices of the characters based on the persona \dots It is not perfect by any means but it is a good starting point.}''
\end{quote}

In both situations, professionals found it beneficial to delegate tasks to GAI without considering the impact on their own skill development.  

% In hindsight, most resisters and embracers implemented strategies within their own quadrants (see Figure \ref{fig:strategiestable}). However, we observed that some individuals (n=6) deviated from this behavior and incorporated multiple strategies across quadrants into their practices. For example, P12, a freelance content strategist who was ambivalent about using GAI technologies in her work, employed both GAI-delegation and human-driven expansion strategies to shape her role. In another example, P18's practices employed both human-driven expansion and delegation strategies in his practices. These two examples indicate a more complex and fluid interaction between professionals' attitudes towards GAI and their chosen crafting approach.

%% file: 5.Discussion.tex
\section{Discussion}
Our findings uncover a nuanced landscape of how writing professionals adapt their practices in response to generative AI (GAI). Here, we compare and contrast resisting and embracing strategies employed by writing professionals to: (1) explore how these strategies shape professionals' creative identity and practices, along with their broader implications for creative work in HCI, (2) generalize findings to broader transformations in GAI-based work, 
%(3) contribute to scholarly discussions on the role of technology in Job Crafting theory, 
and (3) offer concrete recommendations for fostering a more equitable working environment.  

\subsection{\update{Writers' Strategies Around GAI and Its' Broader Implications on Creative Labor}}

\update{Our findings indicate that writing roles with a greater creative focus (e.g., screenwriter (P17), poet (P07)) employed resisting strategies more frequently than technical writing roles (e.g., grant writer (P02), technical writer (P11)), which favored embracing strategies. Also, resistance strategies were more common in better established roles that existed before the digitization of creative work (e.g., author (P18), journalist (P21)), as compared to more emergent roles (e.g., SEO editor (P14), UX writer (P13)).}

\update{Resistance as a rational response is not a new phenomenon in creative work. During the first Industrial Revolution, British textile workers and skilled craftsmen resisted the new weaving machines that they perceived to threaten their work and demean their craftsmanship \cite{Jones2013}. Workers held demonstrations and oversaw the destruction of several machines, aiming to slow down the rapid technological disruption \cite{Roberts2017}. Similar movements have been observed with the introduction of printing and e-books in writing \cite{febvre1997coming}, and photography in art \cite{Marien2006}. However, in contrast to prior literature, the resisting strategies identified in our findings are constructive (i.e., role-enhancing) and multi-dimensional, with professionals who adopted resisting strategies shaping both their \textit{identities} and \textit{practices} to preserve their relevance as craftspeople, thus warranting the label of \textit{adversarial crafting} of their roles in response to GAI. We unpack each of these phenomena in the context of creative labor individually} 

\subsubsection{\update{Shaping Creative Identity: Internal and External Projections}}
One objective of professionals, as part of their resisting strategies, was to preserve and shape their identity through \textit{identity work}, a phenomenon not observed in roles employing embracing strategies. Historically, when workers encounter significant changes such as devaluation, marginalization, and/or discrimination, they react either by protecting their current identity or by restructuring their identities into a desired state as part of their identity work \cite{Petriglieri2011, Ahuja2023}. When technological disruptions are the contributors to these changes, workers either repurpose the technology as an extension of their identity \cite{palani24} or resist the technology to differentiate their own distinct identity \cite{varanasi2019}. Prior HCI literature has shown the latter perceptions to be much stronger in more creative roles, such as artists or creative writers \cite{Van2004, Pierce2001}. 
% For instance, the platformization of creative labor has demonstrated how algorithmic control limits how creative freelancers perceive, construct, and express their identities, pushing workers to devise creative workarounds \cite{munoz22, munoz24}. 
Scholars studying creativity-support tools (CST) are increasingly recognizing this connection, designing explicit pathways within creative tools to enable workers' efforts to build emotional connections and repurpose these tools in line with their identities \cite{palani2022}.

\update{However, GAI differs from previous waves of technology due to its anthropomorphizing features, such as its perceived intelligence in generating creative and analytical content and engaging in context-sensitive conversations \cite{draxler24, nov2023putting}. These features encouraged professional writers, particularly those working in creative domains, to perceive GAI as having its own distinct identity that directly competed with their creative identities. Integrating GAI into their work meant introducing a contrasting identity that risked diluting how they perceived their internal identity, which underpinned their creative outcomes. In response, we observed evidence of activities aimed at reinforcing their sense of worth that took the form of resisting strategies (see Section \ref{identity}). These resisting strategies offered professionals a means of reaffirming their internal identities, in contrast to embracing strategies, which were more commonly employed in roles where such identity conflicts were not as pronounced. These findings extend the preliminary arguments from emergent HCI studies \cite{biermann2022}. Furthermore, while we did not observe direct evidence, we predict that professionals employing a combination of resisting and embracing strategies might exhibit the \textit{AI ghostwriting effect} \cite{draxler24}, whereby they may have a greater tendency to downplay their use of GAI to preserve their creative identities.} 

\update{The professionals who employed resisting strategies also reaffirmed their \textit{external} identity, making it explicitly visible and more distinct. Invisible work, described in the HCI and Science and Technology Studies (STS) literatures as labor that is unacknowledged and undervalued, is particularly likely to emerge in technology-mediated environments \cite{Suchman_1995}. Invisible work can occur when either the worker’s identity or their work is obscured. A common example is articulation work \cite{Star_Strauss_1999}, whereby creative workers engage in coordination and integration tasks essential for achieving broader goals that often go unnoticed and unrecognized \cite{Cheon24, Hatton_2017}. Professionals who invested in resisting strategies made significant efforts to make their identity visible to external stakeholders, including colleagues and end-users (readers of their work). They took deliberate steps to highlight their human involvement at every stage, from job applications to final work delivery (Section \ref{expansion-1}), even in seemingly minor tasks such as articulation work. In contrast, in the case of professionals employing embracing strategies, we did not observe any identity work, indicating that they were more comfortable allowing their stakeholders to perceive a blended version of their identity and GAI's identity.}

\subsubsection{\update{Shaping Creative Practices: Rivalry with GAI or AI Managerial Labor?}}

\update{The perceived differences between individuals' professional identities and the identity of GAI also led them to shape their \textit{practices} in ways that opposed GAI, aiming to preserve their relevance as craftspeople. We found that professionals developed strategies that target GAI's perceived weaknesses, such as its limited ability to generate localized and contextual content (see Section \ref{skillsets-1}), while countering its perceived strengths, such as by bypassing steps in workflow to compete with its ability to generate content rapidly. These behaviors reflect \textit{rivalry}, historically observed among individuals in creative domains \cite{Arora2021}. \citet{Kilduff2010} defines rivalry as ``\textit{a subjective competitive relationship that an actor has with another actor that entails increased psychological involvement and perceived stakes of competition for the focal actor, independent of the objective characteristics of the situation.}'' Interviewees used terms such as \textit{``opposing these tools},'' \textit{at odds with it [GAI]},'' and ``\textit{compete with these tools}'' to describe their strategies for reducing GAI’s influence in their work and profession, while actively competing with it. Interestingly, these practices of rivalry were \textit{motivating}; instead of avoiding GAI, professionals constructed a perceived image of GAI’s capabilities and then crafted their identities and practices to excel in their roles by outperforming it. This form of rivalry mirrors dynamics studied among creative roles, such as creative entrepreneurs \cite{Ahuja2023}, where the contagious nature of passion fuels rivalry. These insights contribute a novel perspective to understanding Human-AI (dis)engagement, a field of literature that has predominantly focused on cooperation and collaboration practices \cite{palani24, vimpari23, ko2023}.}

\update{In contrast, professionals who employed embracing strategies (both GAI-enabled expansion and delegation) used GAI to enhance their practices. This came at a cost; demanding new forms of invisible work through AI managerial labor that remained largely unseen.} As GAI technology advances, this invisible work is likely to only increase, mirroring trends that have been observed in other forms of work exposed to AI, such as the patchwork and repair work necessary to maintain AI systems \cite{fox2023, watkins2023} or in \update{creative work, such as game development \cite{boucher2024} and graphic design \cite{Li24}}. Advances in GAI will also further integrate GAI technologies into workers' processes. In our findings, this was limited to \update{information foraging, creative ideation, overcoming creative roadblocks, and using GAI tools as sounding boards}, but it raises an important question: should the creative contribution of GAI be treated the same as the creative contributions of humans and receive similar recognition? If so, should its work be equally visible? Answers to these questions are essential to define equitable work as the nature of work is fundamentally changing.

\subsection{Contextualizing Findings in the Broader Discussion of GAI and Work}

\subsubsection{GAI and Worker Productivity} 
%1
Previous studies have advocated for GAI in work settings by highlighting productivity and efficiency gains, demonstrating increased worker output \cite{Noy_Zhang_2023, Brynjolfsson2023}. However, our findings suggest that these metrics fail to capture the craftsmanship of writing professionals' work. On the surface, professionals using GAI-driven delegation appeared more productive by offloading time-consuming, repetitive, and stressful tasks (Section \ref{delegate}), while those leveraging GAI-driven expansion employed AI for ideation and content generation (Section \ref{expansion-2-enhancing}). These actions are easily quantifiable in terms of \textit{short-term} efficiency gains. However, the black box nature of these technologies also forced professionals into extensive trial and error to make AI managerial labor effective, increasing the estimation of the labor involved. Our findings suggest that the productivity gains associated with GAI may be offset, or even negated in certain tasks, by the hidden burden of AI managerial labor, a form of invisible labor overlooked in prior studies assessing these technologies \cite{Brynjolfsson2023}.

Metrics such as productivity and efficiency were even less effective in evaluating the performance of professionals who resisted GAI. In some cases, human-driven expansion and localization strategies reduced productivity in the short-term, as these professionals devoted significant time to preserving human labor (e.g., differentiating their writing through human-driven expansion or bypassing steps in workflows to stay competitive in human-driven localization). However, these activities repositioned workers into higher value-added domains that may yield \textit{long-term} gains. Common examples include creating original content or carving out niche roles through localization (section \ref{niche}). This highlights the asymmetries of traditional performance metrics for embracers versus resisters. Traditional performance metrics reward outcomes visible in the short-term, favoring professionals who embrace GAI. They fail to capture the work that requires more time and whose benefits only become apparent later. Developing more comprehensive performance metrics reflecting both immediate and long-term value creation would be both more equitable and better motivate sustained effort and outcomes for both groups while better reflecting the broad societal and professional implications of GAI.

\subsubsection{GAI and Skill Development}
We also observed differences between resisters and embracers in how they developed skills while working with GAI. In GAI-enabled expansion strategies, professionals used GAI for brainstorming and generating alternative ideas (Section \ref{expansion-2-enhancing}). Although they did not always integrate GAI's recommendations into their work, these interactions provided opportunities to expand their knowledge and develop new skills and expertise. This relational method of developing expertise through social interaction used to be thought of as a distinctive feature of human interaction \cite{Anteby2016}. In our findings, GAI was able to take on a relational role, providing context and adapting to professionals' requirements to develop professionals' skills, which had previously only been possible through human collaborations~\cite{Pakarinen2023}.

Professionals who resisted GAI also found ways to develop their skills and expertise. Those who employed human-driven expansion strategies acquired skills that provided greater ownership and control over their work (Section \ref{identity}). Similarly, professionals using human-driven localization strategies focused on refining their skills and tailoring their work to better meet the needs of a niche audience that appreciated what they could offer (Section \ref{niche}). Both of these strategies involved professionals anticipating the likely future implications of GAI-enabled work in their field and taking preemptive action to bring about a more desirable future state, strategies GAI is less effective at.

On the flip side, increased dependence on GAI for more complex and significant tasks showed signs of producing deprofessionalization and skill erosion among participants. Deprofessionalization is defined as the loss of autonomy and control, especially in the context of specialized knowledge or skill requirements in work settings \cite{Haug_1975}. We observed instances of deprofessionalization when professionals using GAI-based delegation strategies outsourced not only boring and mundane tasks but also those tasks that were relevant to professional identity, but challenging to execute (e.g., writing in a specific voice, see Section \ref{imp-tasks}). 

What makes these practices concerning is that, unlike previously reported deprofessionalization due to technology, in which peripheral tasks were delegated to GAI \cite{palani24}, the tasks delegated to it in our study were central to the professionals' identity and delegating them could cause skill decay. One way to mitigate observed instances of deprofessionalization is by designing GAI tools to offer more structured guidance, where the system offers instruction and demonstration on \textit{how} to complete tasks in a modularized approach rather than providing the final solutions in one go. Such an approach would shift the focus from open-ended, outcome-driven interactions to more collaborative, process-oriented engagement, reducing deprofessionalization, and helping users develop and retain their skills. 

\subsubsection{GAI and Job Displacement}
%3
Some of professionals' skill development efforts related to GAI took them beyond their original scope of work. The interviews offered evidence of this role expansion in both professionals who resisted and those who embraced GAI. In human-driven expansion strategies, writing professionals such as journalists learned skills complementary to their role (e.g., graphic design) to broaden their capabilities, with the intention of resisting GAI-enabled automation (Section \ref{reduced-effort}). Similarly, in GAI-enabled delegation strategies, writing professionals such as senior editors drastically reduced, and in some cases eliminated, their dependencies on other roles (e.g., junior writers).

We argue that both of these strategies indicate that roles are expanding in scope, with writing professionals absorbing into their role tasks traditionally executed by others. The key difference between the two groups lies in the direction of skill expansion. Professionals who resisted GAI absorbed the skills of neighboring roles performed by peers at a similar level in the organizational hierarchy whose tasks are complementary to their own. In aggregate, this could lead to role consolidation and redistribution of tasks over time. On the contrary, professionals embracing GAI absorbed the skills of roles that reported to them, which may imply job displacement over time, but in a limited way because there is less incentive to expand the role of experts who can command higher pay to incorporate the responsibilities of their much less well-paid subordinates. These nuanced distinctions offer important contributions to work in economics forecasting employment and industry changes associated with GAI \cite{Autor2023}.  

\subsection{Creating an Equitable Work Environment for Resisters and Embracers of GAI}
Our observations revealed that embracers and resisters were largely operating in silos, unaware of the crafting practices employed by the other group. Future HCI research will benefit from exploring mechanisms to illuminate these distinct practices and foster dialogue that can benefit embracers and resisters. For this, we invoke \citet{Wenger2008}'s notion of Communities of Practice. In a community of practice, individuals with a common professional interest engage in collective learning by sharing lived experiences. Regardless of their particular strategies, embracers and resisters share common interests in safeguarding their jobs against potential negative impacts (e.g., automation) from GAI. Communities of practice could highlight these shared interests, build awareness of GAI's realistic capabilities (for resisters), and identify ways to make professionals' work identity and practices more visible (for embracers). Communities of practice can help both groups move beyond inaccurate assumptions and develop more nuanced mental models of how GAI benefits or hinders their work. The resultant dialogues can also support the development of policies and guidelines that incorporate the practices and perspectives of both groups. 

Lastly, providing both groups with the exposure to the inner workings of GAI's functionalities can further improve how they craft for and against GAI technologies. End-user Explainable AI approaches offer a practical route to achieving this. For our argument, we apply \citet{ehsan2021}'s Explainable AI design, grounded in the notion of social transparency \cite{stuart12}. In this approach, the explainability of the AI system is derived from making the actions of other people visible in the human-AI assemblage \cite{ehsan2021}. Providing explanations for GAI's output, such as the author of the original source used to derive the output, or explaining why the original source was used for the output, can make embracers of GAI more cognizant of how and at what stages they are using the technology. For instance, seeing that the output for a task they delegated is derived from individual authors they admire, rather than thinking of it as output from a "machine," could encourage them to reflect on the types of activities for which they rely on GAI. This is essential, because unlike the discrete nature of interactions with prior forms of AI (e.g., decision making in predictive AI), professionals using GAI were engaging in more dynamic interactions, making it much more difficult for them to psychologically distance themselves to be more reflective and evaluative when using GAI outputs. Explanations may help encourage reflection and distance. Alternatively, for professionals resisting GAI, these explanations can improve their understanding of GAI, alleviate their fears, and further aid in shaping their crafting strategies. Taken together, these approaches can facilitate a more thoughtful and intentional engagement of embracers and resisters with GAI, allowing both groups to maximize its benefits while minimizing potential risks. 

\subsection{Limitations}
Our study had several limitations. Although our study covers a significant cross section of different job roles in writing, our insights are constrained by the general methodological limitations of qualitative research, such as a small sample size and limited generalizability beyond the writing field. More studies in this direction can improve the robustness of these findings across diverse professional fields and populations. Second, our study focused on individuals who had access to GAI and the autonomy to craft their work practices around it, as required by Job Crafting theory. We acknowledge that many professionals, particularly in marginalized communities globally, may lack access to GAI tools or the motivation, agency, or autonomy to craft their own jobs.

%% file: 6.Conclusion.tex
\section{Conclusion}
Our study adds much needed nuance to current understandings of how members of the writing community are shaping their practices by resisting and embracing GAI. It demonstrates that writing professionals are crafting their work, both by resisting as well as embracing GAI. Regardless of their orientation toward GAI, they actively reconfigured their work in response to these technologies through four different crafting strategies.  Those who resisted GAI employed strategies to maximize their human potential, reinforce their professional identity, and carve out their own professional niche. In contrast, those who embraced GAI employed \textit{GAI-enabled} strategies only in specific aspects of their work, such as enhancing workflows and minimizing mundane tasks, while also collaborating with GAI to reduce reliance on colleagues.

\begin{acks}
 This work was supported by NSF grants 1928614 and 2129076. We thank Matthew Law, Rosanna Bellini, and the anonymous reviewers for their thoughtful feedback, which helped refine and improve this paper.
\end{acks}

%% file: 8.appendix.tex
\newpage

\appendix
\twocolumn[
\section{Codebook from Analysis of Interview Data}
\label{app:codebook}
]
\begin{table}[h]
\begin{minipage}[b]{2\linewidth}
 \center
 \renewcommand\arraystretch{1.3}
    \begin{tabular}{|ll|ll|}
        \hline
{\bf Theme} / Code       & Count  &  {\bf Theme} / Code       & Count   \\
        \hline
{\bf Crafting factors around GAI (16\%)} & {\bf 234} & {\bf Changes in task around GAI (29\%)} & {\bf 413} \\
\hline 
Sense of control on work & 13 & Reducing quality/number of tasks & 20 \\
Job characteristics & 14 & Task reconfigurations & 36 \\
Organization push & 15 & Delegating repetitive/tedious tasks & 32 \\
Competition & 19 & Increasing task efficiency & 26 \\
Automation fear & 21 & Work (re)-planning & 19 \\
Skepticism & 11 & New AI-specific tasks & 37 \\
Changing work identity & 19 & Increased effort in low-priority tasks & 07 \\
Role impact & 25 & Increasing visibility of tasks & 35 \\
Competition & 26 & Confirmations & 26 \\
Need to adapt & 12 & Brainstorming ideas & 32 \\
GAI limitations & 11 & Increasing visibility of tasks & 32 \\
Ambiguity & 17 & Upskilling human labor & 37 \\
Environmental changes & 31 & Delegating stressful tasks & 21 \\
& & Increase content ownership & 24\\
& & Niche Appeal & 29 \\
\cline{1-4} 

{\bf Changes in relationship around GAI (17\%)} & {\bf 245} & {\bf Changes in cognition around GAI (16\%)} & {\bf 231} \\
\hline
Increased collaborations w/ AI & 31& Improving internal worth & 26 \\
Peer support to learn AI & 27 & Shifting image/work value & 21 \\
Improving collaborations through AI & 22 & Expanding role & 18 \\
Reducing dependencies & 29  & Improved understanding & 18 \\
Improving identity/brand w/ stakeholders & 22& Reducing emotional stress & 32 \\
Safeguarding community & 12 & Augmenting capabilities & 25 \\
Avoiding GAI promotions & 28 & Improving control & 32 \\
Emotional support & 33  & Competitor to GAI & 24 \\
Instrumental support & 31 & Adversarial reactions & 35 \\
Community mobilization & 10 & & \\
\hline

{\bf GAI specific labor (12\%)} & {\bf 173} & {\bf Levels of GAI integration (10\%)} & {\bf 148} \\
\hline
Prompt curation & 18 & Invisible use & 35 \\
Prompt designing & 31 & Partial use & 26 \\
Refining GAI outputs & 34 & Conformational use & 24 \\
Contributing expertise to improve GAI & 09 & Collaboration & 29 \\
Offloading AI labor & 11 & Modularization & 09 \\
Upskilling GAI techniques/training & 17 & Resistance & 25 \\
Integration/planning negotiations & 24 & & \\
Non-use demand & 29 & & \\
\hline
    \end{tabular}
    \vspace{4pt}
    \caption{The complete codebook that resulted from our analysis of qualitative interviews, showing six themes (bold) and corresponding codes, including the prevalence (\%) for each theme, and the total count for each theme/code.}
    \label{tab:codebook}
    \end{minipage}
\end{table}

%% file: main.bbl
%%% -*-BibTeX-*-
%%% Do NOT edit. File created by BibTeX with style
%%% ACM-Reference-Format-Journals [18-Jan-2012].

\begin{thebibliography}{119}

%%% ====================================================================
%%% NOTE TO THE USER: you can override these defaults by providing
%%% customized versions of any of these macros before the \bibliography
%%% command.  Each of them MUST provide its own final punctuation,
%%% except for \shownote{}, \showDOI{}, and \showURL{}.  The latter two
%%% do not use final punctuation, in order to avoid confusing it with
%%% the Web address.
%%%
%%% To suppress output of a particular field, define its macro to expand
%%% to an empty string, or better, \unskip, like this:
%%%
%%% \newcommand{\showDOI}[1]{\unskip}   % LaTeX syntax
%%%
%%% \def \showDOI #1{\unskip}           % plain TeX syntax
%%%
%%% ====================================================================

\ifx \showCODEN    \undefined \def \showCODEN     #1{\unskip}     \fi
\ifx \showDOI      \undefined \def \showDOI       #1{#1}\fi
\ifx \showISBNx    \undefined \def \showISBNx     #1{\unskip}     \fi
\ifx \showISBNxiii \undefined \def \showISBNxiii  #1{\unskip}     \fi
\ifx \showISSN     \undefined \def \showISSN      #1{\unskip}     \fi
\ifx \showLCCN     \undefined \def \showLCCN      #1{\unskip}     \fi
\ifx \shownote     \undefined \def \shownote      #1{#1}          \fi
\ifx \showarticletitle \undefined \def \showarticletitle #1{#1}   \fi
\ifx \showURL      \undefined \def \showURL       {\relax}        \fi
% The following commands are used for tagged output and should be
% invisible to TeX
\providecommand\bibfield[2]{#2}
\providecommand\bibinfo[2]{#2}
\providecommand\natexlab[1]{#1}
\providecommand\showeprint[2][]{arXiv:#2}

\bibitem[\protect\citeauthoryear{??}{Mar}{2006}]%
        {Marien2006}
 \bibinfo{year}{2006}\natexlab{}.
\newblock \bibinfo{booktitle}{\emph{Photography: a cultural history} (\bibinfo{edition}{2. ed} ed.)}.
\newblock \bibinfo{publisher}{King}, \bibinfo{address}{London}.
\newblock
\showISBNx{9781856694933}


\bibitem[\protect\citeauthoryear{??}{Aro}{2021}]%
        {Arora2021}
 \bibinfo{year}{2021}\natexlab{}.
\newblock \bibinfo{booktitle}{\emph{Competition: what it is and why it happens} (\bibinfo{edition}{first edition.} ed.)}.
\newblock \bibinfo{publisher}{Oxford University Press}, \bibinfo{address}{Oxford; New York, NY}.
\newblock
\showISBNx{9780192652850}


\bibitem[\protect\citeauthoryear{Ahuja}{Ahuja}{2023}]%
        {Ahuja2023}
\bibfield{author}{\bibinfo{person}{Sumati Ahuja}.} \bibinfo{year}{2023}\natexlab{}.
\newblock \showarticletitle{Professional Identity Threats in Interprofessional Collaborations: A Case of Architects in Professional Service Firms}.
\newblock \bibinfo{journal}{\emph{Journal of Management Studies}} \bibinfo{volume}{60}, \bibinfo{number}{2} (\bibinfo{date}{March} \bibinfo{year}{2023}), \bibinfo{pages}{428–453}.
\newblock
\showISSN{0022-2380, 1467-6486}
\urldef\tempurl%
\url{https://doi.org/10.1111/joms.12847}
\showDOI{\tempurl}


\bibitem[\protect\citeauthoryear{Al~Naqbi, Bahroun, and Ahmed}{Al~Naqbi et~al\mbox{.}}{2024}]%
        {Al2024}
\bibfield{author}{\bibinfo{person}{Humaid Al~Naqbi}, \bibinfo{person}{Zied Bahroun}, {and} \bibinfo{person}{Vian Ahmed}.} \bibinfo{year}{2024}\natexlab{}.
\newblock \showarticletitle{Enhancing Work Productivity through Generative Artificial Intelligence: A Comprehensive Literature Review}.
\newblock \bibinfo{journal}{\emph{Sustainability}} \bibinfo{volume}{16}, \bibinfo{number}{3} (\bibinfo{date}{Jan.} \bibinfo{year}{2024}), \bibinfo{pages}{1166}.
\newblock
\showISSN{2071-1050}
\urldef\tempurl%
\url{https://doi.org/10.3390/su16031166}
\showDOI{\tempurl}


\bibitem[\protect\citeauthoryear{Anteby, Chan, and DiBenigno}{Anteby et~al\mbox{.}}{2016}]%
        {Anteby2016}
\bibfield{author}{\bibinfo{person}{Michel Anteby}, \bibinfo{person}{Curtis~K. Chan}, {and} \bibinfo{person}{Julia DiBenigno}.} \bibinfo{year}{2016}\natexlab{}.
\newblock \showarticletitle{Three Lenses on Occupations and Professions in Organizations: Becoming, Doing, and Relating}.
\newblock \bibinfo{journal}{\emph{Academy of Management Annals}} \bibinfo{volume}{10}, \bibinfo{number}{1} (\bibinfo{date}{Jan.} \bibinfo{year}{2016}), \bibinfo{pages}{183–244}.
\newblock
\showISSN{1941-6520, 1941-6067}
\urldef\tempurl%
\url{https://doi.org/10.5465/19416520.2016.1120962}
\showDOI{\tempurl}


\bibitem[\protect\citeauthoryear{Arakawa, Yakura, and Goto}{Arakawa et~al\mbox{.}}{2023}]%
        {arakawa23}
\bibfield{author}{\bibinfo{person}{Riku Arakawa}, \bibinfo{person}{Hiromu Yakura}, {and} \bibinfo{person}{Masataka Goto}.} \bibinfo{year}{2023}\natexlab{}.
\newblock \showarticletitle{CatAlyst: Domain-Extensible Intervention for Preventing Task Procrastination Using Large Generative Models}. In \bibinfo{booktitle}{\emph{Proceedings of the 2023 CHI Conference on Human Factors in Computing Systems}} (Hamburg, Germany) \emph{(\bibinfo{series}{CHI '23})}. \bibinfo{publisher}{Association for Computing Machinery}, \bibinfo{address}{New York, NY, USA}, Article \bibinfo{articleno}{157}, \bibinfo{numpages}{19}~pages.
\newblock
\showISBNx{9781450394215}
\urldef\tempurl%
\url{https://doi.org/10.1145/3544548.3581133}
\showDOI{\tempurl}


\bibitem[\protect\citeauthoryear{Autor, Mindell, Reynolds, and Solow}{Autor et~al\mbox{.}}{2023}]%
        {Autor2023}
\bibfield{author}{\bibinfo{person}{David~H. Autor}, \bibinfo{person}{David~A. Mindell}, \bibinfo{person}{Elisabeth~B. Reynolds}, {and} \bibinfo{person}{Robert~M. Solow}.} \bibinfo{year}{2023}\natexlab{}.
\newblock \bibinfo{booktitle}{\emph{The work of the future: building better jobs in an age of intelligent machines}}.
\newblock \bibinfo{publisher}{MIT Press}, \bibinfo{address}{Cambridge, Massachusetts}.
\newblock
\showISBNx{9780262547307}


\bibitem[\protect\citeauthoryear{Bendoly, Chandrasekaran, Lima, Handfield, Khajavi, and Roscoe}{Bendoly et~al\mbox{.}}{2023}]%
        {Bendoly2023}
\bibfield{author}{\bibinfo{person}{Elliot Bendoly}, \bibinfo{person}{Aravind Chandrasekaran}, \bibinfo{person}{Mateus Do Rego~Ferreira Lima}, \bibinfo{person}{Robert Handfield}, \bibinfo{person}{Siavash~H. Khajavi}, {and} \bibinfo{person}{Samuel Roscoe}.} \bibinfo{year}{2023}\natexlab{}.
\newblock \showarticletitle{The role of generative design and additive manufacturing capabilities in developing human–AI symbiosis: Evidence from multiple case studies}.
\newblock \bibinfo{journal}{\emph{Decision Sciences}} (\bibinfo{date}{Oct.} \bibinfo{year}{2023}), \bibinfo{pages}{deci.12619}.
\newblock
\showISSN{0011-7315, 1540-5915}
\urldef\tempurl%
\url{https://doi.org/10.1111/deci.12619}
\showDOI{\tempurl}


\bibitem[\protect\citeauthoryear{Benharrak, Zindulka, Lehmann, Heuer, and Buschek}{Benharrak et~al\mbox{.}}{2024}]%
        {Benharrak2024}
\bibfield{author}{\bibinfo{person}{Karim Benharrak}, \bibinfo{person}{Tim Zindulka}, \bibinfo{person}{Florian Lehmann}, \bibinfo{person}{Hendrik Heuer}, {and} \bibinfo{person}{Daniel Buschek}.} \bibinfo{year}{2024}\natexlab{}.
\newblock \showarticletitle{Writer-Defined AI Personas for On-Demand Feedback Generation}. In \bibinfo{booktitle}{\emph{Proceedings of the CHI Conference on Human Factors in Computing Systems}} (Honolulu, HI, USA) \emph{(\bibinfo{series}{CHI '24})}. \bibinfo{publisher}{Association for Computing Machinery}, \bibinfo{address}{New York, NY, USA}, Article \bibinfo{articleno}{1049}, \bibinfo{numpages}{18}~pages.
\newblock
\showISBNx{9798400703300}
\urldef\tempurl%
\url{https://doi.org/10.1145/3613904.3642406}
\showDOI{\tempurl}


\bibitem[\protect\citeauthoryear{Berg, Grant, and Johnson}{Berg et~al\mbox{.}}{2010}]%
        {Berg2010}
\bibfield{author}{\bibinfo{person}{Justin~M. Berg}, \bibinfo{person}{Adam~M. Grant}, {and} \bibinfo{person}{Victoria Johnson}.} \bibinfo{year}{2010}\natexlab{}.
\newblock \showarticletitle{When Callings Are Calling: Crafting Work and Leisure in Pursuit of Unanswered Occupational Callings}.
\newblock \bibinfo{journal}{\emph{Organization Science}} \bibinfo{volume}{21}, \bibinfo{number}{5} (\bibinfo{date}{Oct.} \bibinfo{year}{2010}), \bibinfo{pages}{973–994}.
\newblock
\showISSN{1047-7039, 1526-5455}
\urldef\tempurl%
\url{https://doi.org/10.1287/orsc.1090.0497}
\showDOI{\tempurl}


\bibitem[\protect\citeauthoryear{Biermann, Ma, and Yoon}{Biermann et~al\mbox{.}}{2022}]%
        {biermann2022}
\bibfield{author}{\bibinfo{person}{Oloff~C. Biermann}, \bibinfo{person}{Ning~F. Ma}, {and} \bibinfo{person}{Dongwook Yoon}.} \bibinfo{year}{2022}\natexlab{}.
\newblock \showarticletitle{From Tool to Companion: Storywriters Want AI Writers to Respect Their Personal Values and Writing Strategies} \emph{(\bibinfo{series}{DIS '22})}. \bibinfo{publisher}{Association for Computing Machinery}, \bibinfo{address}{New York, NY, USA}, \bibinfo{pages}{1209–1227}.
\newblock
\showISBNx{9781450393584}
\urldef\tempurl%
\url{https://doi.org/10.1145/3532106.3533506}
\showDOI{\tempurl}


\bibitem[\protect\citeauthoryear{Bishop}{Bishop}{2019}]%
        {Bishop_2019}
\bibfield{author}{\bibinfo{person}{Sophie Bishop}.} \bibinfo{year}{2019}\natexlab{}.
\newblock \showarticletitle{Managing visibility on YouTube through algorithmic gossip}.
\newblock \bibinfo{journal}{\emph{New Media \& Society}} \bibinfo{volume}{21}, \bibinfo{number}{11–12} (\bibinfo{date}{Nov.} \bibinfo{year}{2019}), \bibinfo{pages}{2589–2606}.
\newblock
\showISSN{1461-4448, 1461-7315}
\urldef\tempurl%
\url{https://doi.org/10.1177/1461444819854731}
\showDOI{\tempurl}


\bibitem[\protect\citeauthoryear{Bizzi}{Bizzi}{2017}]%
        {Bizzi_2017}
\bibfield{author}{\bibinfo{person}{Lorenzo Bizzi}.} \bibinfo{year}{2017}\natexlab{}.
\newblock \showarticletitle{Network characteristics: When an individual’s job crafting depends on the jobs of others}.
\newblock \bibinfo{journal}{\emph{Human Relations}} \bibinfo{volume}{70}, \bibinfo{number}{4} (\bibinfo{date}{April} \bibinfo{year}{2017}), \bibinfo{pages}{436–460}.
\newblock
\showISSN{0018-7267, 1741-282X}
\urldef\tempurl%
\url{https://doi.org/10.1177/0018726716658963}
\showDOI{\tempurl}


\bibitem[\protect\citeauthoryear{Boucher, Smith, and Telliel}{Boucher et~al\mbox{.}}{2024}]%
        {boucher2024}
\bibfield{author}{\bibinfo{person}{Josiah~D Boucher}, \bibinfo{person}{Gillian Smith}, {and} \bibinfo{person}{Yunus~Do\u{g}an Telliel}.} \bibinfo{year}{2024}\natexlab{}.
\newblock \showarticletitle{Is Resistance Futile?: Early Career Game Developers, Generative AI, and Ethical Skepticism}. In \bibinfo{booktitle}{\emph{Proceedings of the CHI Conference on Human Factors in Computing Systems}} (Honolulu, HI, USA) \emph{(\bibinfo{series}{CHI '24})}. \bibinfo{publisher}{Association for Computing Machinery}, \bibinfo{address}{New York, NY, USA}, Article \bibinfo{articleno}{173}, \bibinfo{numpages}{13}~pages.
\newblock
\showISBNx{9798400703300}
\urldef\tempurl%
\url{https://doi.org/10.1145/3613904.3641889}
\showDOI{\tempurl}


\bibitem[\protect\citeauthoryear{Braun and Clarke}{Braun and Clarke}{2006}]%
        {Braun_Clarke_2006}
\bibfield{author}{\bibinfo{person}{Virginia Braun} {and} \bibinfo{person}{Victoria Clarke}.} \bibinfo{year}{2006}\natexlab{}.
\newblock \showarticletitle{Using thematic analysis in psychology}.
\newblock \bibinfo{journal}{\emph{Qualitative Research in Psychology}} \bibinfo{volume}{3}, \bibinfo{number}{2} (\bibinfo{date}{Jan.} \bibinfo{year}{2006}), \bibinfo{pages}{77–101}.
\newblock
\showISSN{1478-0887, 1478-0895}
\urldef\tempurl%
\url{https://doi.org/10.1191/1478088706qp063oa}
\showDOI{\tempurl}


\bibitem[\protect\citeauthoryear{Bruning and Campion}{Bruning and Campion}{2018}]%
        {Bruning_Campion_2018}
\bibfield{author}{\bibinfo{person}{Patrick~F. Bruning} {and} \bibinfo{person}{Michael~A. Campion}.} \bibinfo{year}{2018}\natexlab{}.
\newblock \showarticletitle{A Role–resource Approach–avoidance Model of Job Crafting: A Multimethod Integration and Extension of Job Crafting Theory}.
\newblock \bibinfo{journal}{\emph{Academy of Management Journal}} \bibinfo{volume}{61}, \bibinfo{number}{2} (\bibinfo{date}{April} \bibinfo{year}{2018}), \bibinfo{pages}{499–522}.
\newblock
\showISSN{0001-4273, 1948-0989}
\urldef\tempurl%
\url{https://doi.org/10.5465/amj.2015.0604}
\showDOI{\tempurl}


\bibitem[\protect\citeauthoryear{Bruning and Campion}{Bruning and Campion}{2019}]%
        {Bruning2019}
\bibfield{author}{\bibinfo{person}{Patrick~F. Bruning} {and} \bibinfo{person}{Michael~A. Campion}.} \bibinfo{year}{2019}\natexlab{}.
\newblock \showarticletitle{Exploring job crafting: Diagnosing and responding to the ways employees adjust their jobs}.
\newblock \bibinfo{journal}{\emph{Business Horizons}} \bibinfo{volume}{62}, \bibinfo{number}{5} (\bibinfo{date}{Sept.} \bibinfo{year}{2019}), \bibinfo{pages}{625–635}.
\newblock
\showISSN{00076813}
\urldef\tempurl%
\url{https://doi.org/10.1016/j.bushor.2019.05.003}
\showDOI{\tempurl}


\bibitem[\protect\citeauthoryear{Brynjolfsson, Li, and Raymond}{Brynjolfsson et~al\mbox{.}}{2023}]%
        {Brynjolfsson2023}
\bibfield{author}{\bibinfo{person}{Erik Brynjolfsson}, \bibinfo{person}{Danielle Li}, {and} \bibinfo{person}{Lindsey Raymond}.} \bibinfo{year}{2023}\natexlab{}.
\newblock \showarticletitle{Generative AI at Work}.
\newblock  \bibinfo{number}{w31161} (\bibinfo{date}{April} \bibinfo{year}{2023}), \bibinfo{pages}{w31161}.
\newblock
\urldef\tempurl%
\url{https://doi.org/10.3386/w31161}
\showDOI{\tempurl}


\bibitem[\protect\citeauthoryear{California State~University, Deng, Joshi, and University}{California State~University et~al\mbox{.}}{2016}]%
        {Dominguez2016}
\bibfield{author}{\bibinfo{person}{Dominguez~Hills California State~University}, \bibinfo{person}{Xuefei Deng}, \bibinfo{person}{K.D. Joshi}, {and} \bibinfo{person}{Washington~State University}.} \bibinfo{year}{2016}\natexlab{}.
\newblock \showarticletitle{Why Individuals Participate in Micro-task Crowdsourcing Work Environment: Revealing Crowdworkers’ Perceptions}.
\newblock \bibinfo{journal}{\emph{Journal of the Association for Information Systems}} \bibinfo{volume}{17}, \bibinfo{number}{10} (\bibinfo{date}{Nov.} \bibinfo{year}{2016}), \bibinfo{pages}{648–673}.
\newblock
\showISSN{15369323}
\urldef\tempurl%
\url{https://doi.org/10.17705/1jais.00441}
\showDOI{\tempurl}


\bibitem[\protect\citeauthoryear{Carr and Hayes}{Carr and Hayes}{2015}]%
        {Carr_Hayes_2015}
\bibfield{author}{\bibinfo{person}{Caleb~T. Carr} {and} \bibinfo{person}{Rebecca~A. Hayes}.} \bibinfo{year}{2015}\natexlab{}.
\newblock \showarticletitle{Social Media: Defining, Developing, and Divining}.
\newblock \bibinfo{journal}{\emph{Atlantic Journal of Communication}} \bibinfo{volume}{23}, \bibinfo{number}{1} (\bibinfo{date}{Jan.} \bibinfo{year}{2015}), \bibinfo{pages}{46–65}.
\newblock
\showISSN{1545-6870, 1545-6889}
\urldef\tempurl%
\url{https://doi.org/10.1080/15456870.2015.972282}
\showDOI{\tempurl}


\bibitem[\protect\citeauthoryear{Chan and Hu}{Chan and Hu}{2023}]%
        {Chan_Hu_2023}
\bibfield{author}{\bibinfo{person}{Cecilia Ka~Yuk Chan} {and} \bibinfo{person}{Wenjie Hu}.} \bibinfo{year}{2023}\natexlab{}.
\newblock \showarticletitle{Students’ voices on generative AI: perceptions, benefits, and challenges in higher education}.
\newblock \bibinfo{journal}{\emph{International Journal of Educational Technology in Higher Education}} \bibinfo{volume}{20}, \bibinfo{number}{1} (\bibinfo{date}{July} \bibinfo{year}{2023}), \bibinfo{pages}{43}.
\newblock
\showISSN{2365-9440}
\urldef\tempurl%
\url{https://doi.org/10.1186/s41239-023-00411-8}
\showDOI{\tempurl}


\bibitem[\protect\citeauthoryear{Cheon and Xu}{Cheon and Xu}{2024}]%
        {Cheon24}
\bibfield{author}{\bibinfo{person}{EunJeong Cheon} {and} \bibinfo{person}{Shengyang Xu}.} \bibinfo{year}{2024}\natexlab{}.
\newblock \showarticletitle{Creative Precarity in Motion: Revealing the Hidden Labor Behind Animating Virtual Characters}. In \bibinfo{booktitle}{\emph{Proceedings of the 2024 ACM Designing Interactive Systems Conference}} (Copenhagen, Denmark) \emph{(\bibinfo{series}{DIS '24})}. \bibinfo{publisher}{Association for Computing Machinery}, \bibinfo{address}{New York, NY, USA}, \bibinfo{pages}{3471–3484}.
\newblock
\showISBNx{9798400705830}
\urldef\tempurl%
\url{https://doi.org/10.1145/3643834.3661545}
\showDOI{\tempurl}


\bibitem[\protect\citeauthoryear{Choi, Kang, Lee, and Kim}{Choi et~al\mbox{.}}{2023}]%
        {choi23}
\bibfield{author}{\bibinfo{person}{Yoonseo Choi}, \bibinfo{person}{Eun~Jeong Kang}, \bibinfo{person}{Min~Kyung Lee}, {and} \bibinfo{person}{Juho Kim}.} \bibinfo{year}{2023}\natexlab{}.
\newblock \showarticletitle{Creator-friendly Algorithms: Behaviors, Challenges, and Design Opportunities in Algorithmic Platforms}. In \bibinfo{booktitle}{\emph{Proceedings of the 2023 CHI Conference on Human Factors in Computing Systems}} (Hamburg, Germany) \emph{(\bibinfo{series}{CHI '23})}. \bibinfo{publisher}{Association for Computing Machinery}, \bibinfo{address}{New York, NY, USA}, Article \bibinfo{articleno}{564}, \bibinfo{numpages}{22}~pages.
\newblock
\showISBNx{9781450394215}
\urldef\tempurl%
\url{https://doi.org/10.1145/3544548.3581386}
\showDOI{\tempurl}


\bibitem[\protect\citeauthoryear{Chung, He, and Adar}{Chung et~al\mbox{.}}{2021}]%
        {chung2021}
\bibfield{author}{\bibinfo{person}{John Joon~Young Chung}, \bibinfo{person}{Shiqing He}, {and} \bibinfo{person}{Eytan Adar}.} \bibinfo{year}{2021}\natexlab{}.
\newblock \showarticletitle{The Intersection of Users, Roles, Interactions, and Technologies in Creativity Support Tools}. In \bibinfo{booktitle}{\emph{Proceedings of the 2021 ACM Designing Interactive Systems Conference}} (Virtual Event, USA) \emph{(\bibinfo{series}{DIS '21})}. \bibinfo{publisher}{Association for Computing Machinery}, \bibinfo{address}{New York, NY, USA}, \bibinfo{pages}{1817–1833}.
\newblock
\showISBNx{9781450384766}
\urldef\tempurl%
\url{https://doi.org/10.1145/3461778.3462050}
\showDOI{\tempurl}


\bibitem[\protect\citeauthoryear{Chung, He, and Adar}{Chung et~al\mbox{.}}{2022}]%
        {chung22}
\bibfield{author}{\bibinfo{person}{John Joon~Young Chung}, \bibinfo{person}{Shiqing He}, {and} \bibinfo{person}{Eytan Adar}.} \bibinfo{year}{2022}\natexlab{}.
\newblock \showarticletitle{Artist Support Networks: Implications for Future Creativity Support Tools}. In \bibinfo{booktitle}{\emph{Proceedings of the 2022 ACM Designing Interactive Systems Conference}} (Virtual Event, Australia) \emph{(\bibinfo{series}{DIS '22})}. \bibinfo{publisher}{Association for Computing Machinery}, \bibinfo{address}{New York, NY, USA}, \bibinfo{pages}{232–246}.
\newblock
\showISBNx{9781450393584}
\urldef\tempurl%
\url{https://doi.org/10.1145/3532106.3533505}
\showDOI{\tempurl}


\bibitem[\protect\citeauthoryear{Creswell and Miller}{Creswell and Miller}{2000}]%
        {Creswell_Miller_2000}
\bibfield{author}{\bibinfo{person}{John~W. Creswell} {and} \bibinfo{person}{Dana~L. Miller}.} \bibinfo{year}{2000}\natexlab{}.
\newblock \showarticletitle{Determining Validity in Qualitative Inquiry}.
\newblock \bibinfo{journal}{\emph{Theory Into Practice}} \bibinfo{volume}{39}, \bibinfo{number}{3} (\bibinfo{date}{Aug.} \bibinfo{year}{2000}), \bibinfo{pages}{124–130}.
\newblock
\showISSN{0040-5841, 1543-0421}
\urldef\tempurl%
\url{https://doi.org/10.1207/s15430421tip3903_2}
\showDOI{\tempurl}


\bibitem[\protect\citeauthoryear{Dell’Acqua, McFowland, Mollick, Lifshitz-Assaf, Kellogg, Rajendran, Krayer, Candelon, and Lakhani}{Dell’Acqua et~al\mbox{.}}{2023}]%
        {Dell2023}
\bibfield{author}{\bibinfo{person}{Fabrizio Dell’Acqua}, \bibinfo{person}{Edward McFowland}, \bibinfo{person}{Ethan~R. Mollick}, \bibinfo{person}{Hila Lifshitz-Assaf}, \bibinfo{person}{Katherine Kellogg}, \bibinfo{person}{Saran Rajendran}, \bibinfo{person}{Lisa Krayer}, \bibinfo{person}{François Candelon}, {and} \bibinfo{person}{Karim~R. Lakhani}.} \bibinfo{year}{2023}\natexlab{}.
\newblock \showarticletitle{Navigating the Jagged Technological Frontier: Field Experimental Evidence of the Effects of AI on Knowledge Worker Productivity and Quality}.
\newblock \bibinfo{journal}{\emph{SSRN Electronic Journal}} (\bibinfo{year}{2023}).
\newblock
\showISSN{1556-5068}
\urldef\tempurl%
\url{https://doi.org/10.2139/ssrn.4573321}
\showDOI{\tempurl}


\bibitem[\protect\citeauthoryear{Dergaa, Chamari, Zmijewski, and Ben~Saad}{Dergaa et~al\mbox{.}}{2023}]%
        {Dergaa2023}
\bibfield{author}{\bibinfo{person}{Ismail Dergaa}, \bibinfo{person}{Karim Chamari}, \bibinfo{person}{Piotr Zmijewski}, {and} \bibinfo{person}{Helmi Ben~Saad}.} \bibinfo{year}{2023}\natexlab{}.
\newblock \showarticletitle{From human writing to artificial intelligence generated text: examining the prospects and potential threats of ChatGPT in academic writing}.
\newblock \bibinfo{journal}{\emph{Biology of Sport}} \bibinfo{volume}{40}, \bibinfo{number}{2} (\bibinfo{year}{2023}), \bibinfo{pages}{615–622}.
\newblock
\showISSN{0860-021X}
\urldef\tempurl%
\url{https://doi.org/10.5114/biolsport.2023.125623}
\showDOI{\tempurl}


\bibitem[\protect\citeauthoryear{Doshi and Hauser}{Doshi and Hauser}{2023}]%
        {Doshi_Hauser_2023}
\bibfield{author}{\bibinfo{person}{Anil~Rajnikant Doshi} {and} \bibinfo{person}{Oliver Hauser}.} \bibinfo{year}{2023}\natexlab{}.
\newblock \showarticletitle{Generative Artificial Intelligence Enhances Creativity}.
\newblock \bibinfo{journal}{\emph{SSRN Electronic Journal}} (\bibinfo{year}{2023}).
\newblock
\showISSN{1556-5068}
\urldef\tempurl%
\url{https://doi.org/10.2139/ssrn.4535536}
\showDOI{\tempurl}


\bibitem[\protect\citeauthoryear{Dourish, Finlay, Sengers, and Wright}{Dourish et~al\mbox{.}}{2004}]%
        {dourish2004}
\bibfield{author}{\bibinfo{person}{Paul Dourish}, \bibinfo{person}{Janet Finlay}, \bibinfo{person}{Phoebe Sengers}, {and} \bibinfo{person}{Peter Wright}.} \bibinfo{year}{2004}\natexlab{}.
\newblock \showarticletitle{Reflective HCI: towards a critical technical practice}. In \bibinfo{booktitle}{\emph{CHI '04 Extended Abstracts on Human Factors in Computing Systems}} (Vienna, Austria) \emph{(\bibinfo{series}{CHI EA '04})}. \bibinfo{publisher}{Association for Computing Machinery}, \bibinfo{address}{New York, NY, USA}, \bibinfo{pages}{1727–1728}.
\newblock
\showISBNx{1581137036}
\urldef\tempurl%
\url{https://doi.org/10.1145/985921.986203}
\showDOI{\tempurl}


\bibitem[\protect\citeauthoryear{Draxler, Werner, Lehmann, Hoppe, Schmidt, Buschek, and Welsch}{Draxler et~al\mbox{.}}{2024}]%
        {draxler24}
\bibfield{author}{\bibinfo{person}{Fiona Draxler}, \bibinfo{person}{Anna Werner}, \bibinfo{person}{Florian Lehmann}, \bibinfo{person}{Matthias Hoppe}, \bibinfo{person}{Albrecht Schmidt}, \bibinfo{person}{Daniel Buschek}, {and} \bibinfo{person}{Robin Welsch}.} \bibinfo{year}{2024}\natexlab{}.
\newblock \showarticletitle{The AI Ghostwriter Effect: When Users do not Perceive Ownership of AI-Generated Text but Self-Declare as Authors}.
\newblock \bibinfo{journal}{\emph{ACM Trans. Comput.-Hum. Interact.}} \bibinfo{volume}{31}, \bibinfo{number}{2}, Article \bibinfo{articleno}{25} (\bibinfo{date}{Feb.} \bibinfo{year}{2024}), \bibinfo{numpages}{40}~pages.
\newblock
\showISSN{1073-0516}
\urldef\tempurl%
\url{https://doi.org/10.1145/3637875}
\showDOI{\tempurl}


\bibitem[\protect\citeauthoryear{Duffy and Meisner}{Duffy and Meisner}{2023}]%
        {Duffy2023}
\bibfield{author}{\bibinfo{person}{Brooke~Erin Duffy} {and} \bibinfo{person}{Colten Meisner}.} \bibinfo{year}{2023}\natexlab{}.
\newblock \showarticletitle{Platform governance at the margins: Social media creators’ experiences with algorithmic (in)visibility}.
\newblock \bibinfo{journal}{\emph{Media, Culture \& Society}} \bibinfo{volume}{45}, \bibinfo{number}{2} (\bibinfo{date}{March} \bibinfo{year}{2023}), \bibinfo{pages}{285–304}.
\newblock
\showISSN{0163-4437, 1460-3675}
\urldef\tempurl%
\url{https://doi.org/10.1177/01634437221111923}
\showDOI{\tempurl}


\bibitem[\protect\citeauthoryear{Duffy, Pinch, Sannon, and Sawey}{Duffy et~al\mbox{.}}{2021}]%
        {Duffy2021}
\bibfield{author}{\bibinfo{person}{Brooke~Erin Duffy}, \bibinfo{person}{Annika Pinch}, \bibinfo{person}{Shruti Sannon}, {and} \bibinfo{person}{Megan Sawey}.} \bibinfo{year}{2021}\natexlab{}.
\newblock \showarticletitle{The Nested Precarities of Creative Labor on Social Media}.
\newblock \bibinfo{journal}{\emph{Social Media + Society}} \bibinfo{volume}{7}, \bibinfo{number}{2} (\bibinfo{date}{April} \bibinfo{year}{2021}), \bibinfo{pages}{20563051211021368}.
\newblock
\showISSN{2056-3051, 2056-3051}
\urldef\tempurl%
\url{https://doi.org/10.1177/20563051211021368}
\showDOI{\tempurl}


\bibitem[\protect\citeauthoryear{Dunn, Shklovski, and Bjørn}{Dunn et~al\mbox{.}}{2024}]%
        {Dunn2024}
\bibfield{author}{\bibinfo{person}{Kellie Dunn}, \bibinfo{person}{Irina Shklovski}, {and} \bibinfo{person}{Pernille Bjørn}.} \bibinfo{year}{2024}\natexlab{}.
\newblock \showarticletitle{What Research through Art can bring to CSCW: exploring ambiguous futures of work}.
\newblock \bibinfo{journal}{\emph{i-com}} \bibinfo{volume}{23}, \bibinfo{number}{1} (\bibinfo{date}{April} \bibinfo{year}{2024}), \bibinfo{pages}{33–55}.
\newblock
\showISSN{2196-6826}
\urldef\tempurl%
\url{https://doi.org/10.1515/icom-2023-0038}
\showDOI{\tempurl}


\bibitem[\protect\citeauthoryear{Ehsan, Liao, Muller, Riedl, and Weisz}{Ehsan et~al\mbox{.}}{2021}]%
        {ehsan2021}
\bibfield{author}{\bibinfo{person}{Upol Ehsan}, \bibinfo{person}{Q.~Vera Liao}, \bibinfo{person}{Michael Muller}, \bibinfo{person}{Mark~O. Riedl}, {and} \bibinfo{person}{Justin~D. Weisz}.} \bibinfo{year}{2021}\natexlab{}.
\newblock \showarticletitle{Expanding Explainability: Towards Social Transparency in AI systems}. In \bibinfo{booktitle}{\emph{Proceedings of the 2021 CHI Conference on Human Factors in Computing Systems}} (Yokohama, Japan) \emph{(\bibinfo{series}{CHI '21})}. \bibinfo{publisher}{Association for Computing Machinery}, \bibinfo{address}{New York, NY, USA}, Article \bibinfo{articleno}{82}, \bibinfo{numpages}{19}~pages.
\newblock
\showISBNx{9781450380966}
\urldef\tempurl%
\url{https://doi.org/10.1145/3411764.3445188}
\showDOI{\tempurl}


\bibitem[\protect\citeauthoryear{Eloundou, Manning, Mishkin, and Rock}{Eloundou et~al\mbox{.}}{2023}]%
        {Eloundou_2023}
\bibfield{author}{\bibinfo{person}{Tyna Eloundou}, \bibinfo{person}{Sam Manning}, \bibinfo{person}{Pamela Mishkin}, {and} \bibinfo{person}{Daniel Rock}.} \bibinfo{year}{2023}\natexlab{}.
\newblock \showarticletitle{GPTs are GPTs: An Early Look at the Labor Market Impact Potential of Large Language Models}.
\newblock  (\bibinfo{year}{2023}).
\newblock
\urldef\tempurl%
\url{https://doi.org/10.48550/ARXIV.2303.10130}
\showDOI{\tempurl}


\bibitem[\protect\citeauthoryear{Epstein, Hertzmann, the Investigators~of Human~Creativity, Akten, Farid, Fjeld, Frank, Groh, Herman, Leach, Mahari, Pentland, Russakovsky, Schroeder, and Smith}{Epstein et~al\mbox{.}}{2023}]%
        {Epstein2023}
\bibfield{author}{\bibinfo{person}{Ziv Epstein}, \bibinfo{person}{Aaron Hertzmann}, \bibinfo{person}{the Investigators~of Human~Creativity}, \bibinfo{person}{Memo Akten}, \bibinfo{person}{Hany Farid}, \bibinfo{person}{Jessica Fjeld}, \bibinfo{person}{Morgan~R. Frank}, \bibinfo{person}{Matthew Groh}, \bibinfo{person}{Laura Herman}, \bibinfo{person}{Neil Leach}, \bibinfo{person}{Robert Mahari}, \bibinfo{person}{Alex~“Sandy” Pentland}, \bibinfo{person}{Olga Russakovsky}, \bibinfo{person}{Hope Schroeder}, {and} \bibinfo{person}{Amy Smith}.} \bibinfo{year}{2023}\natexlab{}.
\newblock \showarticletitle{Art and the science of generative AI}.
\newblock \bibinfo{journal}{\emph{Science}} \bibinfo{volume}{380}, \bibinfo{number}{6650} (\bibinfo{date}{June} \bibinfo{year}{2023}), \bibinfo{pages}{1110–1111}.
\newblock
\showISSN{0036-8075, 1095-9203}
\urldef\tempurl%
\url{https://doi.org/10.1126/science.adh4451}
\showDOI{\tempurl}


\bibitem[\protect\citeauthoryear{Epstein, Schroeder, and Newman}{Epstein et~al\mbox{.}}{2022}]%
        {Epstein_2022}
\bibfield{author}{\bibinfo{person}{Ziv Epstein}, \bibinfo{person}{Hope Schroeder}, {and} \bibinfo{person}{Dava Newman}.} \bibinfo{year}{2022}\natexlab{}.
\newblock \showarticletitle{When happy accidents spark creativity: Bringing collaborative speculation to life with generative AI}.
\newblock  (\bibinfo{year}{2022}).
\newblock
\urldef\tempurl%
\url{https://doi.org/10.48550/ARXIV.2206.00533}
\showDOI{\tempurl}


\bibitem[\protect\citeauthoryear{Febvre and Martin}{Febvre and Martin}{1997}]%
        {febvre1997coming}
\bibfield{author}{\bibinfo{person}{L. Febvre} {and} \bibinfo{person}{H.J. Martin}.} \bibinfo{year}{1997}\natexlab{}.
\newblock \bibinfo{booktitle}{\emph{The Coming of the Book: The Impact of Printing 1450-1800}}.
\newblock \bibinfo{publisher}{Verso}.
\newblock
\showISBNx{9781859841082}
\showLCCN{91114488}
\urldef\tempurl%
\url{https://books.google.com/books?id=9opxcMjv4TUC}
\showURL{%
\tempurl}


\bibitem[\protect\citeauthoryear{Fox, Shorey, Kang, Montiel~Valle, and Rodriguez}{Fox et~al\mbox{.}}{2023}]%
        {fox2023}
\bibfield{author}{\bibinfo{person}{Sarah~E. Fox}, \bibinfo{person}{Samantha Shorey}, \bibinfo{person}{Esther~Y. Kang}, \bibinfo{person}{Dominique Montiel~Valle}, {and} \bibinfo{person}{Estefania Rodriguez}.} \bibinfo{year}{2023}\natexlab{}.
\newblock \showarticletitle{Patchwork: The Hidden, Human Labor of AI Integration within Essential Work}.
\newblock \bibinfo{journal}{\emph{Proc. ACM Hum.-Comput. Interact.}} \bibinfo{volume}{7}, \bibinfo{number}{CSCW1}, Article \bibinfo{articleno}{81} (\bibinfo{date}{apr} \bibinfo{year}{2023}), \bibinfo{numpages}{20}~pages.
\newblock
\urldef\tempurl%
\url{https://doi.org/10.1145/3579514}
\showDOI{\tempurl}


\bibitem[\protect\citeauthoryear{Frich, Biskjaer, MacDonald~Vermeulen, Remy, and Dalsgaard}{Frich et~al\mbox{.}}{2019}]%
        {Frinch19}
\bibfield{author}{\bibinfo{person}{Jonas Frich}, \bibinfo{person}{Michael~Mose Biskjaer}, \bibinfo{person}{Lindsay MacDonald~Vermeulen}, \bibinfo{person}{Christian Remy}, {and} \bibinfo{person}{Peter Dalsgaard}.} \bibinfo{year}{2019}\natexlab{}.
\newblock \showarticletitle{Strategies in Creative Professionals' Use of Digital Tools Across Domains}. In \bibinfo{booktitle}{\emph{Proceedings of the 2019 Conference on Creativity and Cognition}} (San Diego, CA, USA) \emph{(\bibinfo{series}{C\&C '19})}. \bibinfo{publisher}{Association for Computing Machinery}, \bibinfo{address}{New York, NY, USA}, \bibinfo{pages}{210–221}.
\newblock
\showISBNx{9781450359177}
\urldef\tempurl%
\url{https://doi.org/10.1145/3325480.3325494}
\showDOI{\tempurl}


\bibitem[\protect\citeauthoryear{Gordon, Demerouti, Le~Blanc, Bakker, Bipp, and Verhagen}{Gordon et~al\mbox{.}}{2018}]%
        {Gordon2018}
\bibfield{author}{\bibinfo{person}{Heather~J. Gordon}, \bibinfo{person}{Evangelia Demerouti}, \bibinfo{person}{Pascale~M. Le~Blanc}, \bibinfo{person}{Arnold~B. Bakker}, \bibinfo{person}{Tanja Bipp}, {and} \bibinfo{person}{Marc~A.M.T. Verhagen}.} \bibinfo{year}{2018}\natexlab{}.
\newblock \showarticletitle{Individual job redesign: Job crafting interventions in healthcare}.
\newblock \bibinfo{journal}{\emph{Journal of Vocational Behavior}}  \bibinfo{volume}{104} (\bibinfo{date}{Feb.} \bibinfo{year}{2018}), \bibinfo{pages}{98–114}.
\newblock
\showISSN{00018791}
\urldef\tempurl%
\url{https://doi.org/10.1016/j.jvb.2017.07.002}
\showDOI{\tempurl}


\bibitem[\protect\citeauthoryear{Grant-Vallone and Ensher}{Grant-Vallone and Ensher}{2017}]%
        {grant2017re}
\bibfield{author}{\bibinfo{person}{Elisa~J Grant-Vallone} {and} \bibinfo{person}{Ellen~A Ensher}.} \bibinfo{year}{2017}\natexlab{}.
\newblock \showarticletitle{Re-crafting careers for mid-career faculty: A qualitative study}.
\newblock \bibinfo{journal}{\emph{Journal of Higher Education Theory and Practice}} \bibinfo{volume}{17}, \bibinfo{number}{5} (\bibinfo{year}{2017}).
\newblock


\bibitem[\protect\citeauthoryear{Gravador and Teng-Calleja}{Gravador and Teng-Calleja}{2018}]%
        {Gravador2018}
\bibfield{author}{\bibinfo{person}{Luz~Nario Gravador} {and} \bibinfo{person}{Mendiola Teng-Calleja}.} \bibinfo{year}{2018}\natexlab{}.
\newblock \showarticletitle{Work-life balance crafting behaviors: an empirical study}.
\newblock \bibinfo{journal}{\emph{Personnel Review}} \bibinfo{volume}{47}, \bibinfo{number}{4} (\bibinfo{date}{May} \bibinfo{year}{2018}), \bibinfo{pages}{786–804}.
\newblock
\showISSN{0048-3486}
\urldef\tempurl%
\url{https://doi.org/10.1108/PR-05-2016-0112}
\showDOI{\tempurl}


\bibitem[\protect\citeauthoryear{Harrison, Rouse, Fisher, and Amabile}{Harrison et~al\mbox{.}}{2022}]%
        {Harrison2022}
\bibfield{author}{\bibinfo{person}{Spencer Harrison}, \bibinfo{person}{Elizabeth Rouse}, \bibinfo{person}{Colin Fisher}, {and} \bibinfo{person}{Teresa Amabile}.} \bibinfo{year}{2022}\natexlab{}.
\newblock \showarticletitle{The Turn toward Creative Work}.
\newblock \bibinfo{journal}{\emph{Academy of Management Collections}} \bibinfo{volume}{1}, \bibinfo{number}{1} (\bibinfo{date}{Aug.} \bibinfo{year}{2022}).
\newblock
\urldef\tempurl%
\url{https://doi.org/10.5465/amc.2021.0003}
\showDOI{\tempurl}


\bibitem[\protect\citeauthoryear{Hatton}{Hatton}{2017}]%
        {Hatton_2017}
\bibfield{author}{\bibinfo{person}{Erin Hatton}.} \bibinfo{year}{2017}\natexlab{}.
\newblock \showarticletitle{Mechanisms of invisibility: rethinking the concept of invisible work}.
\newblock \bibinfo{journal}{\emph{Work, Employment and Society}} \bibinfo{volume}{31}, \bibinfo{number}{2} (\bibinfo{date}{April} \bibinfo{year}{2017}), \bibinfo{pages}{336–351}.
\newblock
\showISSN{0950-0170, 1469-8722}
\urldef\tempurl%
\url{https://doi.org/10.1177/0950017016674894}
\showDOI{\tempurl}


\bibitem[\protect\citeauthoryear{Haug}{Haug}{1975}]%
        {Haug_1975}
\bibfield{author}{\bibinfo{person}{Marie~R. Haug}.} \bibinfo{year}{1975}\natexlab{}.
\newblock \showarticletitle{The Deprofessionalization of Everyone?}
\newblock \bibinfo{journal}{\emph{Sociological Focus}} \bibinfo{volume}{8}, \bibinfo{number}{3} (\bibinfo{date}{Aug.} \bibinfo{year}{1975}), \bibinfo{pages}{197–213}.
\newblock
\showISSN{0038-0237, 2162-1128}
\urldef\tempurl%
\url{https://doi.org/10.1080/00380237.1975.10570899}
\showDOI{\tempurl}


\bibitem[\protect\citeauthoryear{Hearn}{Hearn}{2020}]%
        {Hearn_2020}
\bibfield{author}{\bibinfo{person}{Greg Hearn}.} \bibinfo{year}{2020}\natexlab{}.
\newblock \bibinfo{booktitle}{\emph{The future of creative work: creativity and digital disruption}}.
\newblock \bibinfo{publisher}{Edward Elgar Publishing}.
\newblock
\showISBNx{9781839101106}
\urldef\tempurl%
\url{https://doi.org/10.4337/9781839101106.00007}
\showDOI{\tempurl}


\bibitem[\protect\citeauthoryear{Hennink, Hutter, and Bailey}{Hennink et~al\mbox{.}}{2019}]%
        {Hennink_2019}
\bibfield{author}{\bibinfo{person}{Monique Hennink}, \bibinfo{person}{Inge Hutter}, {and} \bibinfo{person}{Ajay Bailey}.} \bibinfo{year}{2019}\natexlab{}.
\newblock \bibinfo{booktitle}{\emph{Qualitative research methods} (\bibinfo{edition}{2nd} ed.)}.
\newblock \bibinfo{publisher}{SAGE Publications Ltd}, \bibinfo{address}{Thousand Oaks}.
\newblock
\showISBNx{9781473903913}


\bibitem[\protect\citeauthoryear{Hoque, Mashiat, Ghai, Shelton, Chevalier, Kraus, and Elmqvist}{Hoque et~al\mbox{.}}{2024}]%
        {Hoque2024}
\bibfield{author}{\bibinfo{person}{Md~Naimul Hoque}, \bibinfo{person}{Tasfia Mashiat}, \bibinfo{person}{Bhavya Ghai}, \bibinfo{person}{Cecilia~D. Shelton}, \bibinfo{person}{Fanny Chevalier}, \bibinfo{person}{Kari Kraus}, {and} \bibinfo{person}{Niklas Elmqvist}.} \bibinfo{year}{2024}\natexlab{}.
\newblock \showarticletitle{The HaLLMark Effect: Supporting Provenance and Transparent Use of Large Language Models in Writing with Interactive Visualization}. In \bibinfo{booktitle}{\emph{Proceedings of the CHI Conference on Human Factors in Computing Systems}} (Honolulu, HI, USA) \emph{(\bibinfo{series}{CHI '24})}. \bibinfo{publisher}{Association for Computing Machinery}, \bibinfo{address}{New York, NY, USA}, Article \bibinfo{articleno}{1045}, \bibinfo{numpages}{15}~pages.
\newblock
\showISBNx{9798400703300}
\urldef\tempurl%
\url{https://doi.org/10.1145/3613904.3641895}
\showDOI{\tempurl}


\bibitem[\protect\citeauthoryear{Hsu, Tan, and Stantic}{Hsu et~al\mbox{.}}{2024}]%
        {Hsu_Tan_Stantic_2024}
\bibfield{author}{\bibinfo{person}{Cathy~H.C. Hsu}, \bibinfo{person}{Guoxiong Tan}, {and} \bibinfo{person}{Bela Stantic}.} \bibinfo{year}{2024}\natexlab{}.
\newblock \showarticletitle{A fine-tuned tourism-specific generative AI concept}.
\newblock \bibinfo{journal}{\emph{Annals of Tourism Research}}  \bibinfo{volume}{104} (\bibinfo{date}{Jan.} \bibinfo{year}{2024}), \bibinfo{pages}{103723}.
\newblock
\showISSN{01607383}
\urldef\tempurl%
\url{https://doi.org/10.1016/j.annals.2023.103723}
\showDOI{\tempurl}


\bibitem[\protect\citeauthoryear{Hsueh, Ciolfi~Felice, Alaoui, and Mackay}{Hsueh et~al\mbox{.}}{2024}]%
        {Hsueh24}
\bibfield{author}{\bibinfo{person}{Stacy Hsueh}, \bibinfo{person}{Marianela Ciolfi~Felice}, \bibinfo{person}{Sarah~Fdili Alaoui}, {and} \bibinfo{person}{Wendy~E. Mackay}.} \bibinfo{year}{2024}\natexlab{}.
\newblock \showarticletitle{What Counts as ‘Creative’ Work? Articulating Four Epistemic Positions in Creativity-Oriented HCI Research}. In \bibinfo{booktitle}{\emph{Proceedings of the 2024 CHI Conference on Human Factors in Computing Systems}} (Honolulu, HI, USA) \emph{(\bibinfo{series}{CHI '24})}. \bibinfo{publisher}{Association for Computing Machinery}, \bibinfo{address}{New York, NY, USA}, Article \bibinfo{articleno}{497}, \bibinfo{numpages}{15}~pages.
\newblock
\showISBNx{9798400703300}
\urldef\tempurl%
\url{https://doi.org/10.1145/3613904.3642854}
\showDOI{\tempurl}


\bibitem[\protect\citeauthoryear{Hugill and Yang}{Hugill and Yang}{2013}]%
        {Hugill_Yang_2013}
\bibfield{author}{\bibinfo{person}{Andrew Hugill} {and} \bibinfo{person}{Hongji Yang}.} \bibinfo{year}{2013}\natexlab{}.
\newblock \showarticletitle{The creative turn: new challenges for computing}.
\newblock \bibinfo{journal}{\emph{International Journal of Creative Computing}} \bibinfo{volume}{1}, \bibinfo{number}{1} (\bibinfo{year}{2013}), \bibinfo{pages}{4}.
\newblock
\showISSN{2043-8346, 2043-8354}
\urldef\tempurl%
\url{https://doi.org/10.1504/IJCRC.2013.056934}
\showDOI{\tempurl}


\bibitem[\protect\citeauthoryear{Jiang, Brown, Cheng, Khan, Gupta, Workman, Hanna, Flowers, and Gebru}{Jiang et~al\mbox{.}}{2023}]%
        {23Jiang}
\bibfield{author}{\bibinfo{person}{Harry~H. Jiang}, \bibinfo{person}{Lauren Brown}, \bibinfo{person}{Jessica Cheng}, \bibinfo{person}{Mehtab Khan}, \bibinfo{person}{Abhishek Gupta}, \bibinfo{person}{Deja Workman}, \bibinfo{person}{Alex Hanna}, \bibinfo{person}{Johnathan Flowers}, {and} \bibinfo{person}{Timnit Gebru}.} \bibinfo{year}{2023}\natexlab{}.
\newblock \showarticletitle{AI Art and its Impact on Artists}. In \bibinfo{booktitle}{\emph{Proceedings of the 2023 AAAI/ACM Conference on AI, Ethics, and Society}} (Montr\'{e}al, QC, Canada) \emph{(\bibinfo{series}{AIES '23})}. \bibinfo{publisher}{Association for Computing Machinery}, \bibinfo{address}{New York, NY, USA}, \bibinfo{pages}{363–374}.
\newblock
\showISBNx{9798400702310}
\urldef\tempurl%
\url{https://doi.org/10.1145/3600211.3604681}
\showDOI{\tempurl}


\bibitem[\protect\citeauthoryear{Jochum}{Jochum}{2023}]%
        {Jochum_2023}
\bibfield{author}{\bibinfo{person}{Richard Jochum}.} \bibinfo{year}{2023}\natexlab{}.
\newblock \bibinfo{booktitle}{\emph{Emerging Technologies in Art Education: What If I Don’t Want To?}}
\newblock \bibinfo{publisher}{Springer International Publishing}, \bibinfo{address}{Cham}, \bibinfo{pages}{131–148}.
\newblock
\showISBNx{9783031054754}
\urldef\tempurl%
\url{https://doi.org/10.1007/978-3-031-05476-1_8}
\showDOI{\tempurl}


\bibitem[\protect\citeauthoryear{Jones}{Jones}{2013}]%
        {Jones2013}
\bibfield{author}{\bibinfo{person}{Steven~E. Jones}.} \bibinfo{year}{2013}\natexlab{}.
\newblock \bibinfo{booktitle}{\emph{Against Technology} (\bibinfo{edition}{0} ed.)}.
\newblock \bibinfo{publisher}{Routledge}.
\newblock
\showISBNx{9781135522322}
\urldef\tempurl%
\url{https://doi.org/10.4324/9780203960455}
\showDOI{\tempurl}


\bibitem[\protect\citeauthoryear{Kadoma, Aubin Le~Quere, Fu, Munsch, Metaxa, and Naaman}{Kadoma et~al\mbox{.}}{2024}]%
        {kadoma2024}
\bibfield{author}{\bibinfo{person}{Kowe Kadoma}, \bibinfo{person}{Marianne Aubin Le~Quere}, \bibinfo{person}{Xiyu~Jenny Fu}, \bibinfo{person}{Christin Munsch}, \bibinfo{person}{Dana\"{e} Metaxa}, {and} \bibinfo{person}{Mor Naaman}.} \bibinfo{year}{2024}\natexlab{}.
\newblock \showarticletitle{The Role of Inclusion, Control, and Ownership in Workplace AI-Mediated Communication}. In \bibinfo{booktitle}{\emph{Proceedings of the CHI Conference on Human Factors in Computing Systems}} (Honolulu, HI, USA) \emph{(\bibinfo{series}{CHI '24})}. \bibinfo{publisher}{Association for Computing Machinery}, \bibinfo{address}{New York, NY, USA}, Article \bibinfo{articleno}{1016}, \bibinfo{numpages}{10}~pages.
\newblock
\showISBNx{9798400703300}
\urldef\tempurl%
\url{https://doi.org/10.1145/3613904.3642650}
\showDOI{\tempurl}


\bibitem[\protect\citeauthoryear{Kera}{Kera}{2017}]%
        {Kera_2017}
\bibfield{author}{\bibinfo{person}{Denisa Kera}.} \bibinfo{year}{2017}\natexlab{}.
\newblock \showarticletitle{Science Artisans and Open Science Hardware}.
\newblock \bibinfo{journal}{\emph{Bulletin of Science, Technology \& Society}} \bibinfo{volume}{37}, \bibinfo{number}{2} (\bibinfo{date}{June} \bibinfo{year}{2017}), \bibinfo{pages}{97–111}.
\newblock
\showISSN{0270-4676, 1552-4183}
\urldef\tempurl%
\url{https://doi.org/10.1177/0270467618774978}
\showDOI{\tempurl}


\bibitem[\protect\citeauthoryear{Kidd and Birhane}{Kidd and Birhane}{2023}]%
        {Kidd_Birhane_2023}
\bibfield{author}{\bibinfo{person}{Celeste Kidd} {and} \bibinfo{person}{Abeba Birhane}.} \bibinfo{year}{2023}\natexlab{}.
\newblock \showarticletitle{How AI can distort human beliefs}.
\newblock \bibinfo{journal}{\emph{Science}} \bibinfo{volume}{380}, \bibinfo{number}{6651} (\bibinfo{date}{June} \bibinfo{year}{2023}), \bibinfo{pages}{1222–1223}.
\newblock
\showISSN{0036-8075, 1095-9203}
\urldef\tempurl%
\url{https://doi.org/10.1126/science.adi0248}
\showDOI{\tempurl}


\bibitem[\protect\citeauthoryear{Kilduff, Elfenbein, and Staw}{Kilduff et~al\mbox{.}}{2010}]%
        {Kilduff2010}
\bibfield{author}{\bibinfo{person}{Gavin~J. Kilduff}, \bibinfo{person}{Hillary~Anger Elfenbein}, {and} \bibinfo{person}{Barry~M. Staw}.} \bibinfo{year}{2010}\natexlab{}.
\newblock \showarticletitle{The Psychology of Rivalry: A Relationally Dependent Analysis of Competition}.
\newblock \bibinfo{journal}{\emph{Academy of Management Journal}} \bibinfo{volume}{53}, \bibinfo{number}{5} (\bibinfo{date}{Oct.} \bibinfo{year}{2010}), \bibinfo{pages}{943–969}.
\newblock
\showISSN{0001-4273, 1948-0989}
\urldef\tempurl%
\url{https://doi.org/10.5465/amj.2010.54533171}
\showDOI{\tempurl}


\bibitem[\protect\citeauthoryear{Kim, Han, Adar, Kay, and Chung}{Kim et~al\mbox{.}}{2024}]%
        {kim2024}
\bibfield{author}{\bibinfo{person}{Taewook Kim}, \bibinfo{person}{Hyomin Han}, \bibinfo{person}{Eytan Adar}, \bibinfo{person}{Matthew Kay}, {and} \bibinfo{person}{John Joon~Young Chung}.} \bibinfo{year}{2024}\natexlab{}.
\newblock \showarticletitle{Authors' Values and Attitudes Towards AI-bridged Scalable Personalization of Creative Language Arts}. In \bibinfo{booktitle}{\emph{Proceedings of the CHI Conference on Human Factors in Computing Systems}} (Honolulu, HI, USA) \emph{(\bibinfo{series}{CHI '24})}. \bibinfo{publisher}{Association for Computing Machinery}, \bibinfo{address}{New York, NY, USA}, Article \bibinfo{articleno}{31}, \bibinfo{numpages}{16}~pages.
\newblock
\showISBNx{9798400703300}
\urldef\tempurl%
\url{https://doi.org/10.1145/3613904.3642529}
\showDOI{\tempurl}


\bibitem[\protect\citeauthoryear{{Klee, Miles and Yandoli, Krystie Lee}}{{Klee, Miles and Yandoli, Krystie Lee}}{2024}]%
        {rollingstone2024wgastrike}
\bibfield{author}{\bibinfo{person}{{Klee, Miles and Yandoli, Krystie Lee}}.} \bibinfo{year}{2024}\natexlab{}.
\newblock \bibinfo{title}{Why Striking Hollywood Writers Fear an AI Future}.
\newblock \bibinfo{howpublished}{\url{https://www.rollingstone.com/tv-movies/tv-movie-features/wga-strike-hollywood-writers-ai-artificial-intelligence-chatgpt-screenplay-technology-1234728014/}}.
\newblock
\newblock
\shownote{Accessed: 2024-09-07.}


\bibitem[\protect\citeauthoryear{Ko, Park, Jeon, Jo, Kim, and Seo}{Ko et~al\mbox{.}}{2023}]%
        {ko2023}
\bibfield{author}{\bibinfo{person}{Hyung-Kwon Ko}, \bibinfo{person}{Gwanmo Park}, \bibinfo{person}{Hyeon Jeon}, \bibinfo{person}{Jaemin Jo}, \bibinfo{person}{Juho Kim}, {and} \bibinfo{person}{Jinwook Seo}.} \bibinfo{year}{2023}\natexlab{}.
\newblock \showarticletitle{Large-scale Text-to-Image Generation Models for Visual Artists’ Creative Works}. In \bibinfo{booktitle}{\emph{Proceedings of the 28th International Conference on Intelligent User Interfaces}} (Sydney, NSW, Australia) \emph{(\bibinfo{series}{IUI '23})}. \bibinfo{publisher}{Association for Computing Machinery}, \bibinfo{address}{New York, NY, USA}, \bibinfo{pages}{919–933}.
\newblock
\showISBNx{9798400701061}
\urldef\tempurl%
\url{https://doi.org/10.1145/3581641.3584078}
\showDOI{\tempurl}


\bibitem[\protect\citeauthoryear{Kobiella, Flores~López, Waltenberger, Draxler, and Schmidt}{Kobiella et~al\mbox{.}}{2024}]%
        {Kobiella2024}
\bibfield{author}{\bibinfo{person}{Charlotte Kobiella}, \bibinfo{person}{Yarhy~Said Flores~López}, \bibinfo{person}{Franz Waltenberger}, \bibinfo{person}{Fiona Draxler}, {and} \bibinfo{person}{Albrecht Schmidt}.} \bibinfo{year}{2024}\natexlab{}.
\newblock \showarticletitle{“If the Machine Is As Good As Me, Then What Use Am I?” – How the Use of ChatGPT Changes Young Professionals’ Perception of Productivity and Accomplishment}. In \bibinfo{booktitle}{\emph{Proceedings of the CHI Conference on Human Factors in Computing Systems}} \emph{(\bibinfo{series}{CHI ’24})}. \bibinfo{publisher}{Association for Computing Machinery}, \bibinfo{address}{New York, NY, USA}, \bibinfo{pages}{1–16}.
\newblock
\showISBNx{9798400703300}
\urldef\tempurl%
\url{https://doi.org/10.1145/3613904.3641964}
\showDOI{\tempurl}


\bibitem[\protect\citeauthoryear{Kossek, Piszczek, McAlpine, Hammer, and Burke}{Kossek et~al\mbox{.}}{2016}]%
        {Kossek2016}
\bibfield{author}{\bibinfo{person}{Ellen~Ernst Kossek}, \bibinfo{person}{Matthew~M. Piszczek}, \bibinfo{person}{Kristie~L. McAlpine}, \bibinfo{person}{Leslie~B. Hammer}, {and} \bibinfo{person}{Lisa Burke}.} \bibinfo{year}{2016}\natexlab{}.
\newblock \showarticletitle{Filling the Holes: Work Schedulers As Job Crafters of Employment Practice in Long-Term Health Care}.
\newblock \bibinfo{journal}{\emph{ILR Review}} \bibinfo{volume}{69}, \bibinfo{number}{4} (\bibinfo{date}{Aug.} \bibinfo{year}{2016}), \bibinfo{pages}{961–990}.
\newblock
\showISSN{0019-7939, 2162-271X}
\urldef\tempurl%
\url{https://doi.org/10.1177/0019793916642761}
\showDOI{\tempurl}


\bibitem[\protect\citeauthoryear{Krishna~Kumaran, Shi, and Bailey}{Krishna~Kumaran et~al\mbox{.}}{2021}]%
        {Kumaran21}
\bibfield{author}{\bibinfo{person}{Sneha~R. Krishna~Kumaran}, \bibinfo{person}{Wenxuan~Wendy Shi}, {and} \bibinfo{person}{Brian~P Bailey}.} \bibinfo{year}{2021}\natexlab{}.
\newblock \showarticletitle{Am I Ready to Get Feedback? A Taxonomy of Factors Creators Consider Before Seeking Feedback on In-Progress Creative Work}. In \bibinfo{booktitle}{\emph{Proceedings of the 13th Conference on Creativity and Cognition}} (Virtual Event, Italy) \emph{(\bibinfo{series}{C\&C '21})}. \bibinfo{publisher}{Association for Computing Machinery}, \bibinfo{address}{New York, NY, USA}, Article \bibinfo{articleno}{31}, \bibinfo{numpages}{10}~pages.
\newblock
\showISBNx{9781450383769}
\urldef\tempurl%
\url{https://doi.org/10.1145/3450741.3465255}
\showDOI{\tempurl}


\bibitem[\protect\citeauthoryear{Law and Varanasi}{Law and Varanasi}{2025}]%
        {Varanasi_25}
\bibfield{author}{\bibinfo{person}{Matthew Law} {and} \bibinfo{person}{Rama~Adithya Varanasi}.} \bibinfo{year}{2025}\natexlab{}.
\newblock \bibinfo{booktitle}{\emph{Generative AI \& Changing Work: Systematic Review of Practitioner-led Work Transformations through the Lens of Job Crafting}}.
\newblock \bibinfo{publisher}{Springer International Publishing}, \bibinfo{address}{Cham}.
\newblock
\urldef\tempurl%
\url{https://doi.org/10.48550/arXiv.2502.08854}
\showDOI{\tempurl}


\bibitem[\protect\citeauthoryear{Lazazzara, Tims, and De~Gennaro}{Lazazzara et~al\mbox{.}}{2020}]%
        {Lazazzara_2020}
\bibfield{author}{\bibinfo{person}{Alessandra Lazazzara}, \bibinfo{person}{Maria Tims}, {and} \bibinfo{person}{Davide De~Gennaro}.} \bibinfo{year}{2020}\natexlab{}.
\newblock \showarticletitle{The process of reinventing a job: A meta–synthesis of qualitative job crafting research}.
\newblock \bibinfo{journal}{\emph{Journal of Vocational Behavior}}  \bibinfo{volume}{116} (\bibinfo{date}{Feb.} \bibinfo{year}{2020}), \bibinfo{pages}{103267}.
\newblock
\showISSN{00018791}
\urldef\tempurl%
\url{https://doi.org/10.1016/j.jvb.2019.01.001}
\showDOI{\tempurl}


\bibitem[\protect\citeauthoryear{Lee, Liang, and Yang}{Lee et~al\mbox{.}}{2022}]%
        {lee22}
\bibfield{author}{\bibinfo{person}{Mina Lee}, \bibinfo{person}{Percy Liang}, {and} \bibinfo{person}{Qian Yang}.} \bibinfo{year}{2022}\natexlab{}.
\newblock \showarticletitle{CoAuthor: Designing a Human-AI Collaborative Writing Dataset for Exploring Language Model Capabilities}. In \bibinfo{booktitle}{\emph{Proceedings of the 2022 CHI Conference on Human Factors in Computing Systems}} (New Orleans, LA, USA) \emph{(\bibinfo{series}{CHI '22})}. \bibinfo{publisher}{Association for Computing Machinery}, \bibinfo{address}{New York, NY, USA}, Article \bibinfo{articleno}{388}, \bibinfo{numpages}{19}~pages.
\newblock
\showISBNx{9781450391573}
\urldef\tempurl%
\url{https://doi.org/10.1145/3491102.3502030}
\showDOI{\tempurl}


\bibitem[\protect\citeauthoryear{Li, Cao, Lin, Hou, Zhu, and El~Ali}{Li et~al\mbox{.}}{2024a}]%
        {Li24}
\bibfield{author}{\bibinfo{person}{Jie Li}, \bibinfo{person}{Hancheng Cao}, \bibinfo{person}{Laura Lin}, \bibinfo{person}{Youyang Hou}, \bibinfo{person}{Ruihao Zhu}, {and} \bibinfo{person}{Abdallah El~Ali}.} \bibinfo{year}{2024}\natexlab{a}.
\newblock \showarticletitle{User Experience Design Professionals’ Perceptions of Generative Artificial Intelligence}. In \bibinfo{booktitle}{\emph{Proceedings of the 2024 CHI Conference on Human Factors in Computing Systems}} (Honolulu, HI, USA) \emph{(\bibinfo{series}{CHI '24})}. \bibinfo{publisher}{Association for Computing Machinery}, \bibinfo{address}{New York, NY, USA}, Article \bibinfo{articleno}{381}, \bibinfo{numpages}{18}~pages.
\newblock
\showISBNx{9798400703300}
\urldef\tempurl%
\url{https://doi.org/10.1145/3613904.3642114}
\showDOI{\tempurl}


\bibitem[\protect\citeauthoryear{Li, Liang, Peng, and Yin}{Li et~al\mbox{.}}{2024b}]%
        {Li2024}
\bibfield{author}{\bibinfo{person}{Zhuoyan Li}, \bibinfo{person}{Chen Liang}, \bibinfo{person}{Jing Peng}, {and} \bibinfo{person}{Ming Yin}.} \bibinfo{year}{2024}\natexlab{b}.
\newblock \showarticletitle{The Value, Benefits, and Concerns of Generative AI-Powered Assistance in Writing}. In \bibinfo{booktitle}{\emph{Proceedings of the CHI Conference on Human Factors in Computing Systems}}. \bibinfo{publisher}{ACM}, \bibinfo{address}{Honolulu HI USA}, \bibinfo{pages}{1–25}.
\newblock
\showISBNx{9798400703300}
\urldef\tempurl%
\url{https://doi.org/10.1145/3613904.3642625}
\showDOI{\tempurl}


\bibitem[\protect\citeauthoryear{Martin, Hanrahan, O'Neill, and Gupta}{Martin et~al\mbox{.}}{2014}]%
        {martin14}
\bibfield{author}{\bibinfo{person}{David Martin}, \bibinfo{person}{Benjamin~V. Hanrahan}, \bibinfo{person}{Jacki O'Neill}, {and} \bibinfo{person}{Neha Gupta}.} \bibinfo{year}{2014}\natexlab{}.
\newblock \showarticletitle{Being a turker}. In \bibinfo{booktitle}{\emph{Proceedings of the 17th ACM Conference on Computer Supported Cooperative Work \& Social Computing}} (Baltimore, Maryland, USA) \emph{(\bibinfo{series}{CSCW '14})}. \bibinfo{publisher}{Association for Computing Machinery}, \bibinfo{address}{New York, NY, USA}, \bibinfo{pages}{224–235}.
\newblock
\showISBNx{9781450325400}
\urldef\tempurl%
\url{https://doi.org/10.1145/2531602.2531663}
\showDOI{\tempurl}


\bibitem[\protect\citeauthoryear{{McKinsey \& Company}}{{McKinsey \& Company}}{2024}]%
        {mckinsey_eng}
\bibfield{author}{\bibinfo{person}{{McKinsey \& Company}}.} \bibinfo{year}{2024}\natexlab{}.
\newblock \bibinfo{title}{What is Prompt Engineering?}
\newblock
\newblock
\urldef\tempurl%
\url{https://www.mckinsey.com/featured-insights/mckinsey-explainers/what-is-prompt-engineering}
\showURL{%
\tempurl}
\newblock
\shownote{Accessed: 2024-12-05.}


\bibitem[\protect\citeauthoryear{Mim, Nandi, Khan, Dey, and Ahmed}{Mim et~al\mbox{.}}{2024}]%
        {mim2024}
\bibfield{author}{\bibinfo{person}{Nusrat~Jahan Mim}, \bibinfo{person}{Dipannita Nandi}, \bibinfo{person}{Sadaf~Sumyia Khan}, \bibinfo{person}{Arundhuti Dey}, {and} \bibinfo{person}{Syed~Ishtiaque Ahmed}.} \bibinfo{year}{2024}\natexlab{}.
\newblock \showarticletitle{In-Between Visuals and Visible: The Impacts of Text-to-Image Generative AI Tools on Digital Image-making Practices in the Global South}. In \bibinfo{booktitle}{\emph{Proceedings of the CHI Conference on Human Factors in Computing Systems}} (Honolulu, HI, USA) \emph{(\bibinfo{series}{CHI '24})}. \bibinfo{publisher}{Association for Computing Machinery}, \bibinfo{address}{New York, NY, USA}, Article \bibinfo{articleno}{474}, \bibinfo{numpages}{18}~pages.
\newblock
\showISBNx{9798400703300}
\urldef\tempurl%
\url{https://doi.org/10.1145/3613904.3641951}
\showDOI{\tempurl}


\bibitem[\protect\citeauthoryear{Mirowski, Mathewson, Pittman, and Evans}{Mirowski et~al\mbox{.}}{2023}]%
        {mirowski2023}
\bibfield{author}{\bibinfo{person}{Piotr Mirowski}, \bibinfo{person}{Kory~W. Mathewson}, \bibinfo{person}{Jaylen Pittman}, {and} \bibinfo{person}{Richard Evans}.} \bibinfo{year}{2023}\natexlab{}.
\newblock \showarticletitle{Co-Writing Screenplays and Theatre Scripts with Language Models: Evaluation by Industry Professionals}. In \bibinfo{booktitle}{\emph{Proceedings of the 2023 CHI Conference on Human Factors in Computing Systems}} (Hamburg, Germany) \emph{(\bibinfo{series}{CHI '23})}. \bibinfo{publisher}{Association for Computing Machinery}, \bibinfo{address}{New York, NY, USA}, Article \bibinfo{articleno}{355}, \bibinfo{numpages}{34}~pages.
\newblock
\showISBNx{9781450394215}
\urldef\tempurl%
\url{https://doi.org/10.1145/3544548.3581225}
\showDOI{\tempurl}


\bibitem[\protect\citeauthoryear{Miyazaki, Murayama, Uchiba, An, and Kwak}{Miyazaki et~al\mbox{.}}{2024}]%
        {Miyazaki2024}
\bibfield{author}{\bibinfo{person}{Kunihiro Miyazaki}, \bibinfo{person}{Taichi Murayama}, \bibinfo{person}{Takayuki Uchiba}, \bibinfo{person}{Jisun An}, {and} \bibinfo{person}{Haewoon Kwak}.} \bibinfo{year}{2024}\natexlab{}.
\newblock \showarticletitle{Public perception of generative AI on Twitter: an empirical study based on occupation and usage}.
\newblock \bibinfo{journal}{\emph{EPJ Data Science}} \bibinfo{volume}{13}, \bibinfo{number}{1} (\bibinfo{date}{Jan.} \bibinfo{year}{2024}), \bibinfo{pages}{2}.
\newblock
\showISSN{2193-1127}
\urldef\tempurl%
\url{https://doi.org/10.1140/epjds/s13688-023-00445-y}
\showDOI{\tempurl}


\bibitem[\protect\citeauthoryear{Munoz, Dunn, Sawyer, and Michaels}{Munoz et~al\mbox{.}}{2022}]%
        {munoz22}
\bibfield{author}{\bibinfo{person}{Isabel Munoz}, \bibinfo{person}{Michael Dunn}, \bibinfo{person}{Steve Sawyer}, {and} \bibinfo{person}{Emily Michaels}.} \bibinfo{year}{2022}\natexlab{}.
\newblock \showarticletitle{Platform-mediated Markets, Online Freelance Workers and Deconstructed Identities}.
\newblock \bibinfo{journal}{\emph{Proc. ACM Hum.-Comput. Interact.}} \bibinfo{volume}{6}, \bibinfo{number}{CSCW2}, Article \bibinfo{articleno}{367} (\bibinfo{date}{Nov.} \bibinfo{year}{2022}), \bibinfo{numpages}{24}~pages.
\newblock
\urldef\tempurl%
\url{https://doi.org/10.1145/3555092}
\showDOI{\tempurl}


\bibitem[\protect\citeauthoryear{Munoz, Kim, O'Neil, Dunn, and Sawyer}{Munoz et~al\mbox{.}}{2024}]%
        {munoz24}
\bibfield{author}{\bibinfo{person}{Isabel Munoz}, \bibinfo{person}{Pyeonghwa Kim}, \bibinfo{person}{Clea O'Neil}, \bibinfo{person}{Michael Dunn}, {and} \bibinfo{person}{Steve Sawyer}.} \bibinfo{year}{2024}\natexlab{}.
\newblock \showarticletitle{Platformization of Inequality: Gender and Race in Digital Labor Platforms}.
\newblock \bibinfo{journal}{\emph{Proc. ACM Hum.-Comput. Interact.}} \bibinfo{volume}{8}, \bibinfo{number}{CSCW1}, Article \bibinfo{articleno}{108} (\bibinfo{date}{April} \bibinfo{year}{2024}), \bibinfo{numpages}{22}~pages.
\newblock
\urldef\tempurl%
\url{https://doi.org/10.1145/3637385}
\showDOI{\tempurl}


\bibitem[\protect\citeauthoryear{Nov, Singh, and Mann}{Nov et~al\mbox{.}}{2023}]%
        {nov2023putting}
\bibfield{author}{\bibinfo{person}{Oded Nov}, \bibinfo{person}{Nina Singh}, {and} \bibinfo{person}{Devin Mann}.} \bibinfo{year}{2023}\natexlab{}.
\newblock \showarticletitle{Putting ChatGPT’s medical advice to the (Turing) test: survey study}.
\newblock \bibinfo{journal}{\emph{JMIR Medical Education}}  \bibinfo{volume}{9} (\bibinfo{year}{2023}), \bibinfo{pages}{e46939}.
\newblock


\bibitem[\protect\citeauthoryear{Noy and Zhang}{Noy and Zhang}{2023}]%
        {Noy_Zhang_2023}
\bibfield{author}{\bibinfo{person}{Shakked Noy} {and} \bibinfo{person}{Whitney Zhang}.} \bibinfo{year}{2023}\natexlab{}.
\newblock \showarticletitle{Experimental evidence on the productivity effects of generative artificial intelligence}.
\newblock \bibinfo{journal}{\emph{Science}} \bibinfo{volume}{381}, \bibinfo{number}{6654} (\bibinfo{date}{July} \bibinfo{year}{2023}), \bibinfo{pages}{187–192}.
\newblock
\showISSN{0036-8075, 1095-9203}
\urldef\tempurl%
\url{https://doi.org/10.1126/science.adh2586}
\showDOI{\tempurl}


\bibitem[\protect\citeauthoryear{{OpenAI}}{{OpenAI}}{2024}]%
        {openai_guide}
\bibfield{author}{\bibinfo{person}{{OpenAI}}.} \bibinfo{year}{2024}\natexlab{}.
\newblock \bibinfo{title}{Text Generation Guide}.
\newblock
\newblock
\urldef\tempurl%
\url{https://platform.openai.com/docs/guides/text-generation}
\showURL{%
\tempurl}
\newblock
\shownote{Accessed: 2024-12-05.}


\bibitem[\protect\citeauthoryear{Pakarinen and Huising}{Pakarinen and Huising}{2023}]%
        {Pakarinen2023}
\bibfield{author}{\bibinfo{person}{Pauli Pakarinen} {and} \bibinfo{person}{Ruthanne Huising}.} \bibinfo{year}{2023}\natexlab{}.
\newblock \showarticletitle{Relational Expertise: What Machines Can’t Know}.
\newblock \bibinfo{journal}{\emph{Journal of Management Studies}} (\bibinfo{date}{March} \bibinfo{year}{2023}), \bibinfo{pages}{joms.12915}.
\newblock
\showISSN{0022-2380, 1467-6486}
\urldef\tempurl%
\url{https://doi.org/10.1111/joms.12915}
\showDOI{\tempurl}


\bibitem[\protect\citeauthoryear{Palani, Ledo, Fitzmaurice, and Anderson}{Palani et~al\mbox{.}}{2022}]%
        {palani2022}
\bibfield{author}{\bibinfo{person}{Srishti Palani}, \bibinfo{person}{David Ledo}, \bibinfo{person}{George Fitzmaurice}, {and} \bibinfo{person}{Fraser Anderson}.} \bibinfo{year}{2022}\natexlab{}.
\newblock \showarticletitle{``I don’t want to feel like I’m working in a 1960s factory”: The Practitioner Perspective on Creativity Support Tool Adoption}. In \bibinfo{booktitle}{\emph{Proceedings of the 2022 CHI Conference on Human Factors in Computing Systems}} (New Orleans, LA, USA) \emph{(\bibinfo{series}{CHI '22})}. \bibinfo{publisher}{Association for Computing Machinery}, \bibinfo{address}{New York, NY, USA}, Article \bibinfo{articleno}{379}, \bibinfo{numpages}{18}~pages.
\newblock
\showISBNx{9781450391573}
\urldef\tempurl%
\url{https://doi.org/10.1145/3491102.3501933}
\showDOI{\tempurl}


\bibitem[\protect\citeauthoryear{Palani and Ramos}{Palani and Ramos}{2024}]%
        {palani24}
\bibfield{author}{\bibinfo{person}{Srishti Palani} {and} \bibinfo{person}{Gonzalo Ramos}.} \bibinfo{year}{2024}\natexlab{}.
\newblock \showarticletitle{Evolving Roles and Workflows of Creative Practitioners in the Age of Generative AI}. In \bibinfo{booktitle}{\emph{Proceedings of the 16th Conference on Creativity \& Cognition}} (Chicago, IL, USA) \emph{(\bibinfo{series}{C\&C '24})}. \bibinfo{publisher}{Association for Computing Machinery}, \bibinfo{address}{New York, NY, USA}, \bibinfo{pages}{170–184}.
\newblock
\showISBNx{9798400704857}
\urldef\tempurl%
\url{https://doi.org/10.1145/3635636.3656190}
\showDOI{\tempurl}


\bibitem[\protect\citeauthoryear{Petriglieri}{Petriglieri}{2011}]%
        {Petriglieri2011}
\bibfield{author}{\bibinfo{person}{Jennifer~Louise Petriglieri}.} \bibinfo{year}{2011}\natexlab{}.
\newblock \showarticletitle{Under Threat: Responses to and the Consequences of Threats to Individuals’ Identities}.
\newblock \bibinfo{journal}{\emph{Academy of Management Review}} \bibinfo{volume}{36}, \bibinfo{number}{4} (\bibinfo{date}{Oct.} \bibinfo{year}{2011}), \bibinfo{pages}{641–662}.
\newblock
\showISSN{0363-7425, 1930-3807}
\urldef\tempurl%
\url{https://doi.org/10.5465/amr.2009.0087}
\showDOI{\tempurl}


\bibitem[\protect\citeauthoryear{Petrou, Bakker, and Van Den~Heuvel}{Petrou et~al\mbox{.}}{2017}]%
        {Petrou2017}
\bibfield{author}{\bibinfo{person}{Paraskevas Petrou}, \bibinfo{person}{Arnold~B. Bakker}, {and} \bibinfo{person}{Machteld Van Den~Heuvel}.} \bibinfo{year}{2017}\natexlab{}.
\newblock \showarticletitle{Weekly job crafting and leisure crafting: Implications for meaning‐making and work engagement}.
\newblock \bibinfo{journal}{\emph{Journal of Occupational and Organizational Psychology}} \bibinfo{volume}{90}, \bibinfo{number}{2} (\bibinfo{date}{June} \bibinfo{year}{2017}), \bibinfo{pages}{129–152}.
\newblock
\showISSN{0963-1798, 2044-8325}
\urldef\tempurl%
\url{https://doi.org/10.1111/joop.12160}
\showDOI{\tempurl}


\bibitem[\protect\citeauthoryear{Pierce, Kostova, and Dirks}{Pierce et~al\mbox{.}}{2001}]%
        {Pierce2001}
\bibfield{author}{\bibinfo{person}{Jon~L. Pierce}, \bibinfo{person}{Tatiana Kostova}, {and} \bibinfo{person}{Kurt~T. Dirks}.} \bibinfo{year}{2001}\natexlab{}.
\newblock \showarticletitle{Toward a Theory of Psychological Ownership in Organizations}.
\newblock \bibinfo{journal}{\emph{The Academy of Management Review}} \bibinfo{volume}{26}, \bibinfo{number}{2} (\bibinfo{date}{April} \bibinfo{year}{2001}), \bibinfo{pages}{298}.
\newblock
\showISSN{03637425}
\urldef\tempurl%
\url{https://doi.org/10.2307/259124}
\showDOI{\tempurl}


\bibitem[\protect\citeauthoryear{Razaq, Kolko, and Hsieh}{Razaq et~al\mbox{.}}{2022}]%
        {razaq22}
\bibfield{author}{\bibinfo{person}{Lubna Razaq}, \bibinfo{person}{Beth Kolko}, {and} \bibinfo{person}{Gary Hsieh}.} \bibinfo{year}{2022}\natexlab{}.
\newblock \showarticletitle{Making crafting visible while rendering labor invisible on the Etsy platform}. In \bibinfo{booktitle}{\emph{Proceedings of the 2022 ACM Designing Interactive Systems Conference}} (Virtual Event, Australia) \emph{(\bibinfo{series}{DIS '22})}. \bibinfo{publisher}{Association for Computing Machinery}, \bibinfo{address}{New York, NY, USA}, \bibinfo{pages}{424–438}.
\newblock
\showISBNx{9781450393584}
\urldef\tempurl%
\url{https://doi.org/10.1145/3532106.3533573}
\showDOI{\tempurl}


\bibitem[\protect\citeauthoryear{Reza, Laundry, Musabirov, Dushniku, Yu, Mittal, Grossman, Liut, Kuzminykh, and Williams}{Reza et~al\mbox{.}}{2024}]%
        {Reza2024}
\bibfield{author}{\bibinfo{person}{Mohi Reza}, \bibinfo{person}{Nathan~M Laundry}, \bibinfo{person}{Ilya Musabirov}, \bibinfo{person}{Peter Dushniku}, \bibinfo{person}{Zhi Yuan~“Michael” Yu}, \bibinfo{person}{Kashish Mittal}, \bibinfo{person}{Tovi Grossman}, \bibinfo{person}{Michael Liut}, \bibinfo{person}{Anastasia Kuzminykh}, {and} \bibinfo{person}{Joseph~Jay Williams}.} \bibinfo{year}{2024}\natexlab{}.
\newblock \showarticletitle{ABScribe: Rapid Exploration \& Organization of Multiple Writing Variations in Human-AI Co-Writing Tasks using Large Language Models}. In \bibinfo{booktitle}{\emph{Proceedings of the CHI Conference on Human Factors in Computing Systems}} (Honolulu, HI, USA) \emph{(\bibinfo{series}{CHI '24})}. \bibinfo{publisher}{Association for Computing Machinery}, \bibinfo{address}{New York, NY, USA}, Article \bibinfo{articleno}{1042}, \bibinfo{numpages}{18}~pages.
\newblock
\showISBNx{9798400703300}
\urldef\tempurl%
\url{https://doi.org/10.1145/3613904.3641899}
\showDOI{\tempurl}


\bibitem[\protect\citeauthoryear{Roberts}{Roberts}{2017}]%
        {Roberts2017}
\bibfield{author}{\bibinfo{person}{Matthew Roberts}.} \bibinfo{year}{2017}\natexlab{}.
\newblock \showarticletitle{Rural Luddism and the makeshift economy of the Nottinghamshire framework knitters}.
\newblock \bibinfo{journal}{\emph{Social History}} \bibinfo{volume}{42}, \bibinfo{number}{3} (\bibinfo{date}{July} \bibinfo{year}{2017}), \bibinfo{pages}{365–398}.
\newblock
\showISSN{0307-1022, 1470-1200}
\urldef\tempurl%
\url{https://doi.org/10.1080/03071022.2017.1327644}
\showDOI{\tempurl}


\bibitem[\protect\citeauthoryear{Robertson, Walther, and Radcliffe}{Robertson et~al\mbox{.}}{2007}]%
        {Robertson2007}
\bibfield{author}{\bibinfo{person}{Brett~F. Robertson}, \bibinfo{person}{Joachim Walther}, {and} \bibinfo{person}{David~F. Radcliffe}.} \bibinfo{year}{2007}\natexlab{}.
\newblock \showarticletitle{Creativity and the Use of CAD Tools: Lessons for Engineering Design Education From Industry}.
\newblock \bibinfo{journal}{\emph{Journal of Mechanical Design}} \bibinfo{volume}{129}, \bibinfo{number}{7} (\bibinfo{date}{July} \bibinfo{year}{2007}), \bibinfo{pages}{753–760}.
\newblock
\showISSN{1050-0472, 1528-9001}
\urldef\tempurl%
\url{https://doi.org/10.1115/1.2722329}
\showDOI{\tempurl}


\bibitem[\protect\citeauthoryear{Ross}{Ross}{2008}]%
        {Ross_2008}
\bibfield{author}{\bibinfo{person}{Andrew Ross}.} \bibinfo{year}{2008}\natexlab{}.
\newblock \showarticletitle{The New Geography of Work: Power to the Precarious?}
\newblock \bibinfo{journal}{\emph{Theory, Culture \& Society}} \bibinfo{volume}{25}, \bibinfo{number}{7–8} (\bibinfo{date}{Dec.} \bibinfo{year}{2008}), \bibinfo{pages}{31–49}.
\newblock
\showISSN{0263-2764, 1460-3616}
\urldef\tempurl%
\url{https://doi.org/10.1177/0263276408097795}
\showDOI{\tempurl}


\bibitem[\protect\citeauthoryear{Rosselli Del~Turco and Dalsgaard}{Rosselli Del~Turco and Dalsgaard}{2023}]%
        {turco23}
\bibfield{author}{\bibinfo{person}{Emilia Rosselli Del~Turco} {and} \bibinfo{person}{Peter Dalsgaard}.} \bibinfo{year}{2023}\natexlab{}.
\newblock \showarticletitle{"I wouldn’t dare losing one": How music artists capture and manage ideas}. In \bibinfo{booktitle}{\emph{Proceedings of the 15th Conference on Creativity and Cognition}} (Virtual Event, USA) \emph{(\bibinfo{series}{C\&C '23})}. \bibinfo{publisher}{Association for Computing Machinery}, \bibinfo{address}{New York, NY, USA}, \bibinfo{pages}{88–102}.
\newblock
\showISBNx{9798400701801}
\urldef\tempurl%
\url{https://doi.org/10.1145/3591196.3593338}
\showDOI{\tempurl}


\bibitem[\protect\citeauthoryear{Samuelson}{Samuelson}{2023}]%
        {Samuelson_2023}
\bibfield{author}{\bibinfo{person}{Pamela Samuelson}.} \bibinfo{year}{2023}\natexlab{}.
\newblock \showarticletitle{Generative AI meets copyright}.
\newblock \bibinfo{journal}{\emph{Science}} \bibinfo{volume}{381}, \bibinfo{number}{6654} (\bibinfo{date}{July} \bibinfo{year}{2023}), \bibinfo{pages}{158–161}.
\newblock
\showISSN{0036-8075, 1095-9203}
\urldef\tempurl%
\url{https://doi.org/10.1126/science.adi0656}
\showDOI{\tempurl}


\bibitem[\protect\citeauthoryear{Sanchez}{Sanchez}{2023}]%
        {sanchez23}
\bibfield{author}{\bibinfo{person}{T\'{e}o Sanchez}.} \bibinfo{year}{2023}\natexlab{}.
\newblock \showarticletitle{Examining the Text-to-Image Community of Practice: Why and How do People Prompt Generative AIs?}. In \bibinfo{booktitle}{\emph{Proceedings of the 15th Conference on Creativity and Cognition}} (Virtual Event, USA) \emph{(\bibinfo{series}{C\&C '23})}. \bibinfo{publisher}{Association for Computing Machinery}, \bibinfo{address}{New York, NY, USA}, \bibinfo{pages}{43–61}.
\newblock
\showISBNx{9798400701801}
\urldef\tempurl%
\url{https://doi.org/10.1145/3591196.3593051}
\showDOI{\tempurl}


\bibitem[\protect\citeauthoryear{Simpson and Semaan}{Simpson and Semaan}{2023}]%
        {semaan23}
\bibfield{author}{\bibinfo{person}{Ellen Simpson} {and} \bibinfo{person}{Bryan Semaan}.} \bibinfo{year}{2023}\natexlab{}.
\newblock \showarticletitle{Rethinking Creative Labor: A Sociotechnical Examination of Creativity \& Creative Work on TikTok}. In \bibinfo{booktitle}{\emph{Proceedings of the 2023 CHI Conference on Human Factors in Computing Systems}} (Hamburg, Germany) \emph{(\bibinfo{series}{CHI '23})}. \bibinfo{publisher}{Association for Computing Machinery}, \bibinfo{address}{New York, NY, USA}, Article \bibinfo{articleno}{244}, \bibinfo{numpages}{16}~pages.
\newblock
\showISBNx{9781450394215}
\urldef\tempurl%
\url{https://doi.org/10.1145/3544548.3580649}
\showDOI{\tempurl}


\bibitem[\protect\citeauthoryear{Small, Wiesenfeld, Brandfield-Harvey, Jonassen, Mandal, Stevens, Major, Lostraglio, Szerencsy, Jones, Aphinyanaphongs, Johnson, Nov, and Mann}{Small et~al\mbox{.}}{2024}]%
        {Small2024}
\bibfield{author}{\bibinfo{person}{William~R. Small}, \bibinfo{person}{Batia Wiesenfeld}, \bibinfo{person}{Beatrix Brandfield-Harvey}, \bibinfo{person}{Zoe Jonassen}, \bibinfo{person}{Soumik Mandal}, \bibinfo{person}{Elizabeth~R. Stevens}, \bibinfo{person}{Vincent~J. Major}, \bibinfo{person}{Erin Lostraglio}, \bibinfo{person}{Adam Szerencsy}, \bibinfo{person}{Simon Jones}, \bibinfo{person}{Yindalon Aphinyanaphongs}, \bibinfo{person}{Stephen~B. Johnson}, \bibinfo{person}{Oded Nov}, {and} \bibinfo{person}{Devin Mann}.} \bibinfo{year}{2024}\natexlab{}.
\newblock \showarticletitle{Large Language Model–Based Responses to Patients’ In-Basket Messages}.
\newblock \bibinfo{journal}{\emph{JAMA Network Open}} \bibinfo{volume}{7}, \bibinfo{number}{7} (\bibinfo{date}{July} \bibinfo{year}{2024}), \bibinfo{pages}{e2422399}.
\newblock
\showISSN{2574-3805}
\urldef\tempurl%
\url{https://doi.org/10.1001/jamanetworkopen.2024.22399}
\showDOI{\tempurl}


\bibitem[\protect\citeauthoryear{Star and Strauss}{Star and Strauss}{1999}]%
        {Star_Strauss_1999}
\bibfield{author}{\bibinfo{person}{Susan~Leigh Star} {and} \bibinfo{person}{Anselm Strauss}.} \bibinfo{year}{1999}\natexlab{}.
\newblock \showarticletitle{Layers of Silence, Arenas of Voice: The Ecology of Visible and Invisible Work}.
\newblock \bibinfo{journal}{\emph{Computer Supported Cooperative Work (CSCW)}} \bibinfo{volume}{8}, \bibinfo{number}{1–2} (\bibinfo{date}{March} \bibinfo{year}{1999}), \bibinfo{pages}{9–30}.
\newblock
\showISSN{0925-9724, 1573-7551}
\urldef\tempurl%
\url{https://doi.org/10.1023/A:1008651105359}
\showDOI{\tempurl}


\bibitem[\protect\citeauthoryear{Stuart, Dabbish, Kiesler, Kinnaird, and Kang}{Stuart et~al\mbox{.}}{2012}]%
        {stuart12}
\bibfield{author}{\bibinfo{person}{H.~Colleen Stuart}, \bibinfo{person}{Laura Dabbish}, \bibinfo{person}{Sara Kiesler}, \bibinfo{person}{Peter Kinnaird}, {and} \bibinfo{person}{Ruogu Kang}.} \bibinfo{year}{2012}\natexlab{}.
\newblock \showarticletitle{Social transparency in networked information exchange: a theoretical framework}. In \bibinfo{booktitle}{\emph{Proceedings of the ACM 2012 Conference on Computer Supported Cooperative Work}} (Seattle, Washington, USA) \emph{(\bibinfo{series}{CSCW '12})}. \bibinfo{publisher}{Association for Computing Machinery}, \bibinfo{address}{New York, NY, USA}, \bibinfo{pages}{451–460}.
\newblock
\showISBNx{9781450310864}
\urldef\tempurl%
\url{https://doi.org/10.1145/2145204.2145275}
\showDOI{\tempurl}


\bibitem[\protect\citeauthoryear{Suchman}{Suchman}{1995}]%
        {Suchman_1995}
\bibfield{author}{\bibinfo{person}{Lucy Suchman}.} \bibinfo{year}{1995}\natexlab{}.
\newblock \showarticletitle{Making work visible}.
\newblock \bibinfo{journal}{\emph{Commun. ACM}} \bibinfo{volume}{38}, \bibinfo{number}{9} (\bibinfo{date}{Sept.} \bibinfo{year}{1995}), \bibinfo{pages}{56–64}.
\newblock
\showISSN{0001-0782, 1557-7317}
\urldef\tempurl%
\url{https://doi.org/10.1145/223248.223263}
\showDOI{\tempurl}


\bibitem[\protect\citeauthoryear{Suh, Dangol, Meadan, Miller, and Kientz}{Suh et~al\mbox{.}}{2024}]%
        {Suh2024}
\bibfield{author}{\bibinfo{person}{Hyewon Suh}, \bibinfo{person}{Aayushi Dangol}, \bibinfo{person}{Hedda Meadan}, \bibinfo{person}{Carol~A. Miller}, {and} \bibinfo{person}{Julie~A. Kientz}.} \bibinfo{year}{2024}\natexlab{}.
\newblock \showarticletitle{Opportunities and Challenges for AI-Based Support for Speech-Language Pathologists}. In \bibinfo{booktitle}{\emph{Proceedings of the 3rd Annual Meeting of the Symposium on Human-Computer Interaction for Work}}. \bibinfo{publisher}{ACM}, \bibinfo{address}{Newcastle upon Tyne United Kingdom}, \bibinfo{pages}{1–14}.
\newblock
\showISBNx{9798400710179}
\urldef\tempurl%
\url{https://doi.org/10.1145/3663384.3663387}
\showDOI{\tempurl}


\bibitem[\protect\citeauthoryear{Takaffoli, Li, and M\"{a}kel\"{a}}{Takaffoli et~al\mbox{.}}{2024}]%
        {takaffoli}
\bibfield{author}{\bibinfo{person}{Macy Takaffoli}, \bibinfo{person}{Sijia Li}, {and} \bibinfo{person}{Ville M\"{a}kel\"{a}}.} \bibinfo{year}{2024}\natexlab{}.
\newblock \showarticletitle{Generative AI in User Experience Design and Research: How Do UX Practitioners, Teams, and Companies Use GenAI in Industry?}. In \bibinfo{booktitle}{\emph{Proceedings of the 2024 ACM Designing Interactive Systems Conference}} (Copenhagen, Denmark) \emph{(\bibinfo{series}{DIS '24})}. \bibinfo{publisher}{Association for Computing Machinery}, \bibinfo{address}{New York, NY, USA}, \bibinfo{pages}{1579–1593}.
\newblock
\showISBNx{9798400705830}
\urldef\tempurl%
\url{https://doi.org/10.1145/3643834.3660720}
\showDOI{\tempurl}


\bibitem[\protect\citeauthoryear{Timmermans and Tavory}{Timmermans and Tavory}{2012}]%
        {Timmermans_2012}
\bibfield{author}{\bibinfo{person}{Stefan Timmermans} {and} \bibinfo{person}{Iddo Tavory}.} \bibinfo{year}{2012}\natexlab{}.
\newblock \showarticletitle{Theory Construction in Qualitative Research: From Grounded Theory to Abductive Analysis}.
\newblock   \bibinfo{volume}{30} (\bibinfo{date}{Sept.} \bibinfo{year}{2012}), \bibinfo{pages}{167–186}.
\newblock
\showISSN{0735-2751, 1467-9558}
\urldef\tempurl%
\url{https://doi.org/10.1177/0735275112457914}
\showDOI{\tempurl}


\bibitem[\protect\citeauthoryear{Tims, Bakker, and Derks}{Tims et~al\mbox{.}}{2012}]%
        {Tims_2012}
\bibfield{author}{\bibinfo{person}{Maria Tims}, \bibinfo{person}{Arnold~B. Bakker}, {and} \bibinfo{person}{Daantje Derks}.} \bibinfo{year}{2012}\natexlab{}.
\newblock \showarticletitle{Development and validation of the job crafting scale}.
\newblock \bibinfo{journal}{\emph{Journal of Vocational Behavior}} \bibinfo{volume}{80}, \bibinfo{number}{1} (\bibinfo{date}{Feb.} \bibinfo{year}{2012}), \bibinfo{pages}{173–186}.
\newblock
\showISSN{00018791}
\urldef\tempurl%
\url{https://doi.org/10.1016/j.jvb.2011.05.009}
\showDOI{\tempurl}


\bibitem[\protect\citeauthoryear{Tims, Bakker, and Derks}{Tims et~al\mbox{.}}{2013}]%
        {Tims2013}
\bibfield{author}{\bibinfo{person}{Maria Tims}, \bibinfo{person}{Arnold~B. Bakker}, {and} \bibinfo{person}{Daantje Derks}.} \bibinfo{year}{2013}\natexlab{}.
\newblock \showarticletitle{The impact of job crafting on job demands, job resources, and well-being.}
\newblock \bibinfo{journal}{\emph{Journal of Occupational Health Psychology}} \bibinfo{volume}{18}, \bibinfo{number}{2} (\bibinfo{year}{2013}), \bibinfo{pages}{230–240}.
\newblock
\showISSN{1939-1307, 1076-8998}
\urldef\tempurl%
\url{https://doi.org/10.1037/a0032141}
\showDOI{\tempurl}


\bibitem[\protect\citeauthoryear{Van~Dyne and Pierce}{Van~Dyne and Pierce}{2004}]%
        {Van2004}
\bibfield{author}{\bibinfo{person}{Linn Van~Dyne} {and} \bibinfo{person}{Jon~L. Pierce}.} \bibinfo{year}{2004}\natexlab{}.
\newblock \showarticletitle{Psychological ownership and feelings of possession: three field studies predicting employee attitudes and organizational citizenship behavior}.
\newblock \bibinfo{journal}{\emph{Journal of Organizational Behavior}} \bibinfo{volume}{25}, \bibinfo{number}{4} (\bibinfo{date}{June} \bibinfo{year}{2004}), \bibinfo{pages}{439–459}.
\newblock
\showISSN{0894-3796, 1099-1379}
\urldef\tempurl%
\url{https://doi.org/10.1002/job.249}
\showDOI{\tempurl}


\bibitem[\protect\citeauthoryear{Varanasi, Kizilcec, and Dell}{Varanasi et~al\mbox{.}}{2019}]%
        {varanasi2019}
\bibfield{author}{\bibinfo{person}{Rama~Adithya Varanasi}, \bibinfo{person}{Ren\'{e}~F. Kizilcec}, {and} \bibinfo{person}{Nicola Dell}.} \bibinfo{year}{2019}\natexlab{}.
\newblock \showarticletitle{How Teachers in India Reconfigure their Work Practices around a Teacher-Oriented Technology Intervention}.
\newblock \bibinfo{journal}{\emph{Proc. ACM Hum.-Comput. Interact.}} \bibinfo{volume}{3}, \bibinfo{number}{CSCW}, Article \bibinfo{articleno}{220} (\bibinfo{date}{nov} \bibinfo{year}{2019}), \bibinfo{numpages}{21}~pages.
\newblock
\urldef\tempurl%
\url{https://doi.org/10.1145/3359322}
\showDOI{\tempurl}


\bibitem[\protect\citeauthoryear{Vimpari, Kultima, H\"{a}m\"{a}l\"{a}inen, and Guckelsberger}{Vimpari et~al\mbox{.}}{2023}]%
        {vimpari23}
\bibfield{author}{\bibinfo{person}{Veera Vimpari}, \bibinfo{person}{Annakaisa Kultima}, \bibinfo{person}{Perttu H\"{a}m\"{a}l\"{a}inen}, {and} \bibinfo{person}{Christian Guckelsberger}.} \bibinfo{year}{2023}\natexlab{}.
\newblock \showarticletitle{“An Adapt-or-Die Type of Situation”: Perception, Adoption, and Use of Text-to-Image-Generation AI by Game Industry Professionals}.
\newblock \bibinfo{journal}{\emph{Proc. ACM Hum.-Comput. Interact.}} \bibinfo{volume}{7}, \bibinfo{number}{CHI PLAY}, Article \bibinfo{articleno}{379} (\bibinfo{date}{oct} \bibinfo{year}{2023}), \bibinfo{numpages}{34}~pages.
\newblock
\urldef\tempurl%
\url{https://doi.org/10.1145/3611025}
\showDOI{\tempurl}


\bibitem[\protect\citeauthoryear{Walczak and Cellary}{Walczak and Cellary}{2023}]%
        {Walczak_2023}
\bibfield{author}{\bibinfo{person}{Krzysztof Walczak} {and} \bibinfo{person}{Wojciech Cellary}.} \bibinfo{year}{2023}\natexlab{}.
\newblock \showarticletitle{Challenges for higher education in the era of widespread access to generative AI}.
\newblock \bibinfo{journal}{\emph{Economics and Business Review}} \bibinfo{volume}{9}, \bibinfo{number}{2} (\bibinfo{date}{July} \bibinfo{year}{2023}).
\newblock
\showISSN{2450-0097, 2392-1641}
\urldef\tempurl%
\url{https://doi.org/10.18559/ebr.2023.2.743}
\showDOI{\tempurl}


\bibitem[\protect\citeauthoryear{Wang, Menon, Long, Henderson, Li, Crowston, Hansen, Nickerson, and Chilton}{Wang et~al\mbox{.}}{2024}]%
        {wang24}
\bibfield{author}{\bibinfo{person}{Sitong Wang}, \bibinfo{person}{Samia Menon}, \bibinfo{person}{Tao Long}, \bibinfo{person}{Keren Henderson}, \bibinfo{person}{Dingzeyu Li}, \bibinfo{person}{Kevin Crowston}, \bibinfo{person}{Mark Hansen}, \bibinfo{person}{Jeffrey~V Nickerson}, {and} \bibinfo{person}{Lydia~B Chilton}.} \bibinfo{year}{2024}\natexlab{}.
\newblock \showarticletitle{ReelFramer: Human-AI Co-Creation for News-to-Video Translation}. In \bibinfo{booktitle}{\emph{Proceedings of the CHI Conference on Human Factors in Computing Systems}} (Honolulu, HI, USA) \emph{(\bibinfo{series}{CHI '24})}. \bibinfo{publisher}{Association for Computing Machinery}, \bibinfo{address}{New York, NY, USA}, Article \bibinfo{articleno}{169}, \bibinfo{numpages}{20}~pages.
\newblock
\showISBNx{9798400703300}
\urldef\tempurl%
\url{https://doi.org/10.1145/3613904.3642868}
\showDOI{\tempurl}


\bibitem[\protect\citeauthoryear{Washington}{Washington}{2023}]%
        {Washington_2023}
\bibfield{author}{\bibinfo{person}{Jerry Washington}.} \bibinfo{year}{2023}\natexlab{}.
\newblock \showarticletitle{The Impact of Generative Artificial Intelligence on Writer’s Self-Efficacy: A Critical Literature Review}.
\newblock \bibinfo{journal}{\emph{SSRN Electronic Journal}} (\bibinfo{year}{2023}).
\newblock
\showISSN{1556-5068}
\urldef\tempurl%
\url{https://doi.org/10.2139/ssrn.4538043}
\showDOI{\tempurl}


\bibitem[\protect\citeauthoryear{Watkins}{Watkins}{2023}]%
        {watkins2023}
\bibfield{author}{\bibinfo{person}{Elizabeth~Anne Watkins}.} \bibinfo{year}{2023}\natexlab{}.
\newblock \showarticletitle{Face Work: A Human-Centered Investigation into Facial Verification in Gig Work}.
\newblock \bibinfo{journal}{\emph{Proc. ACM Hum.-Comput. Interact.}} \bibinfo{volume}{7}, \bibinfo{number}{CSCW1}, Article \bibinfo{articleno}{52} (\bibinfo{date}{apr} \bibinfo{year}{2023}), \bibinfo{numpages}{24}~pages.
\newblock
\urldef\tempurl%
\url{https://doi.org/10.1145/3579485}
\showDOI{\tempurl}


\bibitem[\protect\citeauthoryear{Wenger-Trayner}{Wenger-Trayner}{2008}]%
        {Wenger2008}
\bibfield{author}{\bibinfo{person}{Étienne Wenger-Trayner}.} \bibinfo{year}{2008}\natexlab{}.
\newblock \bibinfo{booktitle}{\emph{Communities of practice: learning, meaning, and identity} (\bibinfo{edition}{18th printing} ed.)}.
\newblock \bibinfo{publisher}{Cambridge University Press}, \bibinfo{address}{Cambridge}.
\newblock
\showISBNx{9780521663632}


\bibitem[\protect\citeauthoryear{Wong, Kost, and Fieseler}{Wong et~al\mbox{.}}{2021}]%
        {Wong_Fieseler_2021}
\bibfield{author}{\bibinfo{person}{Sut~I Wong}, \bibinfo{person}{Dominique Kost}, {and} \bibinfo{person}{Christian Fieseler}.} \bibinfo{year}{2021}\natexlab{}.
\newblock \showarticletitle{From crafting what you do to building resilience for career commitment in the gig economy}.
\newblock \bibinfo{journal}{\emph{Human Resource Management Journal}} \bibinfo{volume}{31}, \bibinfo{number}{4} (\bibinfo{date}{Nov.} \bibinfo{year}{2021}), \bibinfo{pages}{918–935}.
\newblock
\showISSN{0954-5395, 1748-8583}
\urldef\tempurl%
\url{https://doi.org/10.1111/1748-8583.12342}
\showDOI{\tempurl}


\bibitem[\protect\citeauthoryear{Woodruff, Shelby, Kelley, Rousso-Schindler, Smith-Loud, and Wilcox}{Woodruff et~al\mbox{.}}{2024}]%
        {wilcox24}
\bibfield{author}{\bibinfo{person}{Allison Woodruff}, \bibinfo{person}{Renee Shelby}, \bibinfo{person}{Patrick~Gage Kelley}, \bibinfo{person}{Steven Rousso-Schindler}, \bibinfo{person}{Jamila Smith-Loud}, {and} \bibinfo{person}{Lauren Wilcox}.} \bibinfo{year}{2024}\natexlab{}.
\newblock \showarticletitle{How Knowledge Workers Think Generative AI Will (Not) Transform Their Industries}. In \bibinfo{booktitle}{\emph{Proceedings of the CHI Conference on Human Factors in Computing Systems}} (Honolulu, HI, USA) \emph{(\bibinfo{series}{CHI '24})}. \bibinfo{publisher}{Association for Computing Machinery}, \bibinfo{address}{New York, NY, USA}, Article \bibinfo{articleno}{641}, \bibinfo{numpages}{26}~pages.
\newblock
\showISBNx{9798400703300}
\urldef\tempurl%
\url{https://doi.org/10.1145/3613904.3642700}
\showDOI{\tempurl}


\bibitem[\protect\citeauthoryear{Wrzesniewski and Dutton}{Wrzesniewski and Dutton}{2001}]%
        {Wrzesniewski_2001}
\bibfield{author}{\bibinfo{person}{Amy Wrzesniewski} {and} \bibinfo{person}{Jane~E. Dutton}.} \bibinfo{year}{2001}\natexlab{}.
\newblock \showarticletitle{Crafting a Job: Revisioning Employees as Active Crafters of Their Work}.
\newblock \bibinfo{journal}{\emph{The Academy of Management Review}} \bibinfo{volume}{26}, \bibinfo{number}{2} (\bibinfo{date}{April} \bibinfo{year}{2001}), \bibinfo{pages}{179}.
\newblock
\showISSN{03637425}
\urldef\tempurl%
\url{https://doi.org/10.2307/259118}
\showDOI{\tempurl}


\bibitem[\protect\citeauthoryear{Zhang and Parker}{Zhang and Parker}{2019}]%
        {zhang_2019}
\bibfield{author}{\bibinfo{person}{Fangfang Zhang} {and} \bibinfo{person}{Sharon~K. Parker}.} \bibinfo{year}{2019}\natexlab{}.
\newblock \showarticletitle{Reorienting job crafting research: A hierarchical structure of job crafting concepts and integrative review}.
\newblock \bibinfo{journal}{\emph{Journal of Organizational Behavior}} \bibinfo{volume}{40}, \bibinfo{number}{2} (\bibinfo{date}{Feb.} \bibinfo{year}{2019}), \bibinfo{pages}{126–146}.
\newblock
\showISSN{0894-3796, 1099-1379}
\urldef\tempurl%
\url{https://doi.org/10.1002/job.2332}
\showDOI{\tempurl}


\bibitem[\protect\citeauthoryear{Zhou and Sterman}{Zhou and Sterman}{2024}]%
        {zhou2024}
\bibfield{author}{\bibinfo{person}{David Zhou} {and} \bibinfo{person}{Sarah Sterman}.} \bibinfo{year}{2024}\natexlab{}.
\newblock \showarticletitle{Ai.llude: Investigating Rewriting AI-Generated Text to Support Creative Expression}. In \bibinfo{booktitle}{\emph{Proceedings of the 16th Conference on Creativity \& Cognition}} (Chicago, IL, USA) \emph{(\bibinfo{series}{C\&C '24})}. \bibinfo{publisher}{Association for Computing Machinery}, \bibinfo{address}{New York, NY, USA}, \bibinfo{pages}{241–254}.
\newblock
\showISBNx{9798400704857}
\urldef\tempurl%
\url{https://doi.org/10.1145/3635636.3656187}
\showDOI{\tempurl}


\bibitem[\protect\citeauthoryear{Zoran, Shilkrot, Nanyakkara, and Paradiso}{Zoran et~al\mbox{.}}{2014}]%
        {zoran2014}
\bibfield{author}{\bibinfo{person}{Amit Zoran}, \bibinfo{person}{Roy Shilkrot}, \bibinfo{person}{Suranga Nanyakkara}, {and} \bibinfo{person}{Joseph Paradiso}.} \bibinfo{year}{2014}\natexlab{}.
\newblock \showarticletitle{The Hybrid Artisans: A Case Study in Smart Tools}.
\newblock  \bibinfo{volume}{21}, \bibinfo{number}{3}, Article \bibinfo{articleno}{15} (\bibinfo{date}{jun} \bibinfo{year}{2014}), \bibinfo{numpages}{29}~pages.
\newblock
\showISSN{1073-0516}
\urldef\tempurl%
\url{https://doi.org/10.1145/2617570}
\showDOI{\tempurl}


\end{thebibliography}
